\documentclass[12]{article}
\usepackage[table]{xcolor}
\usepackage[utf8]{inputenc}
\usepackage{amsmath}
\usepackage{url}
\usepackage{multirow}
\usepackage{cancel}
\usepackage{array}
\usepackage{varwidth}
\usepackage{tabularx}
\usepackage{tabulary}
\usepackage{wrapfig}
\usepackage{newtxtext,newtxmath}
\usepackage[nottoc,notlot,notlof]{tocbibind}
\usepackage[flushleft]{threeparttable}
\usepackage{tikz,pgfplots,pgfplotstable}

\pgfplotsset{width=\columnwidth,compat=1.14}
\definecolor{LightCyan}{rgb}{0.88,1,1}
\definecolor{LightYellow}{rgb}{1,1,0.5}
\definecolor{LightBlue}{rgb}{0.5,1,1}
\definecolor{LightPurple}{rgb}{1,0.75,1}

\title{Augmented Reality Chess Analyzer (ARChessAnalyzer): In-Device Inference of Physical Chess Game Positions through Board Segmentation and Piece Recognition using Convolutional Neural Networks}
\author{Anav Mehta \\ Cupertino High School \\ Cupertino, CA, USA}

\begin{document}

\maketitle
\begin{abstract}
    Chess game position analysis is important in improving ones game. It requires entry of moves into a chess engine which is, cumbersome and error prone. We present ARChessAnalyzer, a complete pipeline from live image capture of a physical chess game, to board and piece recognition, to move analysis and finally to Augmented Reality (AR) overlay of the chess diagram position and move on the physical board. ARChessAnalyzer is like a scene analyzer - it uses an ensemble of traditional image and vision techniques to segment the scene (ie the chess game) and uses Convolution Neural Networks (CNNs)  to predict the segmented pieces and combine it together to analyze the game. This paper advances the state of the art in the first of its kind end to end integration of robust detection and segmentation of the board, chess piece detection using the fine-tuned AlexNet CNN and chess engine analyzer in a handheld device app.  The accuracy of the entire chess position prediction pipeline is 93.45\% and takes 3-4.5sec from live capture to AR overlay. We also validated our hypothesis that ARChessAnalyzer, is faster at analysis than manual entry for all board positions for valid outcomes. Our hope is that the instantaneous feedback this app provides will help chess learners worldwide at all levels improve their game.
\end{abstract}
\thispagestyle{empty}
\newpage
\thispagestyle{empty}

\section{Introduction}
For a player to improve ones chess game, its important to record the moves, so that you can analyze the game later perhaps in the chess club either with a coach or using an online chess engine. The recording of moves during a chess game is a tedious manual task that is error prone and impedes the flow of the game and also hinders efficient use of time. Instead, imagine if, you can capture a live image of the game, detect the board and all the pieces, predicting the board position and analyze it to provide immediate feedback of the best move available to the player. In an age when much analysis and storage of chess
games is done on computers, chess players at all levels can
benefit from the ability to analyze a game immediately by taking a picture of a real-life board, as opposed to manual input. This paper describes exactly that novel approach of identifying the chess position and next move through the simple means of live picture. This advances several technologies board and piece detection and ties them together on a mobile device to generate a position for a chess engine to analyze the best best move to display on an augmented reality overlay.

Advances in powerful algorithms and hardware computing units, have allowed for deep learning algorithms to solve a wide variety of tasks which were previously deemed difficult for computers to tackle. Challenging problems such as playing strategic games like Go and poker, and visual object recognition \cite{imagenet} are now possible using modern compute environments. A type of artificial neural network, called a Convolutional Neural Network (CNN), has demonstrated capabilities for highly accurate image classification after being trained on a large data set of samples \cite{deep}.

In the first part of the detection pipeline, a simple binary image classifier is built to detect a chessboard, and image and vision techniques from OpenCV library are used to segment the board. 
The popular deep CNN architecture, AlexNet \cite{alexnetfig} - the winner of the 2012 ImageNet Large Scale Visual Recognition Competition, is fine-tuned with a data set of chess pieces. This model is used for piece recognition from the segmented board. The output of the piece detector - a a Forsyth–Edwards Notation(FEN) \cite{fen} position string - is used by the popular chess engine Stockfish \cite{stockfish} for analysis, and finally the best move from the engine is https://www.overleaf.com/project/5e850192a900400001e484ecoverlayed along with the chess diagram on the physical board.  This ensemble of algorithms is integrated in an iOS mobile app in that is an augmented reality chess analysis engine. The app provides immediate feedback and helping chess players with their game.

The remainder of this paper is organized as follows. We will first cover the previous research work and then discuss the implementation and our unique contributions of the different pieces of ARChessAnalyzer pipeline. Finally, we present the app development details, discuss the experimental setup, results and conclusions.

\newpage
\section{Previous Work}
Chess position detection can be broadly separated into two areas. The first, board recognition and segmentation, refers to the detection of the chess board within the image and segmentation of the board into 64 images.  This is a prerequisite to the second step - piece recognition which is a classification of those images - as empty or any of the 12 categories of chess pieces. 

\subsection{Chessboard Recognition and Segmentation}

Chessboard detection has traditionally been solved in the context of camera calibration using hints such as marking the corners at the expense of its versatility. There have also been dedicated hardware attempts for chessboard detection \cite{koray}. 
Recent computer vision advances have led to general techniques for board recognition can be separated into corner-based or line-based approaches \cite{duda}.
Corner-based approaches identify the corners of the chess board, then either perform Hough transforms to identify the lines or assign coordinates to the corners directly . These approaches either assume a plain background, so as to reduce the number of corner-generating artifacts; use a top down overhead view; or require the absence of pieces from the board, in order to prevent the occlusion of corners or lattice points by pieces.
Line-based approaches use edge detection on the input image to identify lines of the chess board. Domain knowledge, such as the fact that the board can be identified by 18 total lines and the orientations of half those lines will be orthogonal to the other half, makes line-based approaches more robust to noise, and is therefore the more popular technique \cite{tam}.  Some proposed corner based approaches which averaged the quality of results while some have utilized exclusively line-based methods \cite{kanchibail}.   \cite{tam} introduced the classification of such methods into line-based and corner-based approaches but were limited to recognizing chessboards without chess pieces on them.

\subsection{Chess Piece Recognition}
Initial attempts at piece recognition were game-tracking applications assume the starting positions of the pieces, and there can use differences in intensity values after each move to track the movement of pieces \cite{koray}.
Techniques that do not assume a starting position focus on color segmentation to detect pieces and then use shape descriptors to identify them. However, color segmentation often relies on square and piece color combinations, so unreasonable constraints are place on type of chessboard  or a sideview that relies on depth but occludes most of the pieces.
 This paper uses a more robust piece recognition solution using trained object detection CNN models. 

\newpage
\section{Materials and Methods}

\subsection{Position Detection Pipeline}
\begin{figure}[h]
  \caption{Position detection pipeline}
  \label{fig:DetectionPipeline}
  \includegraphics[width=\linewidth]{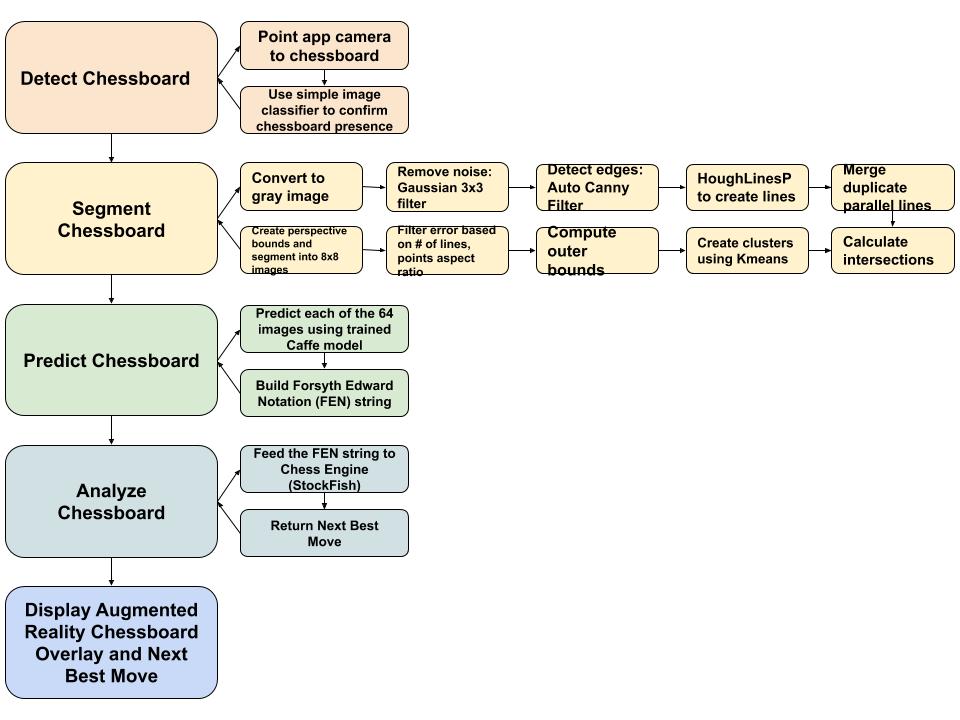}
\end{figure}
We have designed an ensemble of methods and tools with engineering tradeoffs, while advancing the state of the art to develop the entire chess position pipeline in an iOS app (Figure \ref{fig:DetectionPipeline}).  The following are the major steps. 
\begin{itemize}
    \item \textbf{Detect Chessboard:}
        This is a precursor to the entire pipeline, determines the presence of a chessboard using simple binary image classifier.
    \item \textbf{Segment Chessboard:}
        OpenCV image and vision techniques are used to determine the outer bounds of the chessboard and segment the image into 64 pieces.
    \item \textbf{Predict Chessboard:}
        An pretrained CNN object detector model is used to predict the 64 image and form the position string.
    \item \textbf{Analyze Chessboard:}
        The string is fed into an open source chess engine - StockFish to determine the next best move.
    \item \textbf{Augment Reality OverLay Chessboard}
        The next move and position are overlayed on top of the existing chessboard for the player.
\end{itemize}

\subsection{Detecting and Segmenting Chessboard}

\subsubsection{Detecting Chessboard}
In order to begin predicting the board, it is necessary to detect a chessboard. In order to begin detecting, we developed a simple binary image classifier trained using 95 images with a 80/20\% training/test split. We provide user feedback for allowing the user to stay stable while the object is detected. This took less than 1sec from start of the pointing of the board. 

\subsubsection{Segmenting Chessboard}

The algorithm consists of five main steps (Figure \ref{fig:seg_pipe}).  
\begin{itemize}
\item \textbf{Detecting straight and parallel lines}
Besides standard line detector, the additional objective is to merge all small segments that are nearly colinear into long straight lines and then identify the parallel lines.

\begin{figure}[h!]
    \centering
    \caption{Segmentation pipeline}
    \begin{tabular}{|c|c|c|} \hline
    \rowcolor{LightCyan}
    Orig & Gray &  Blur \\ \hline
    \includegraphics[width=30mm]{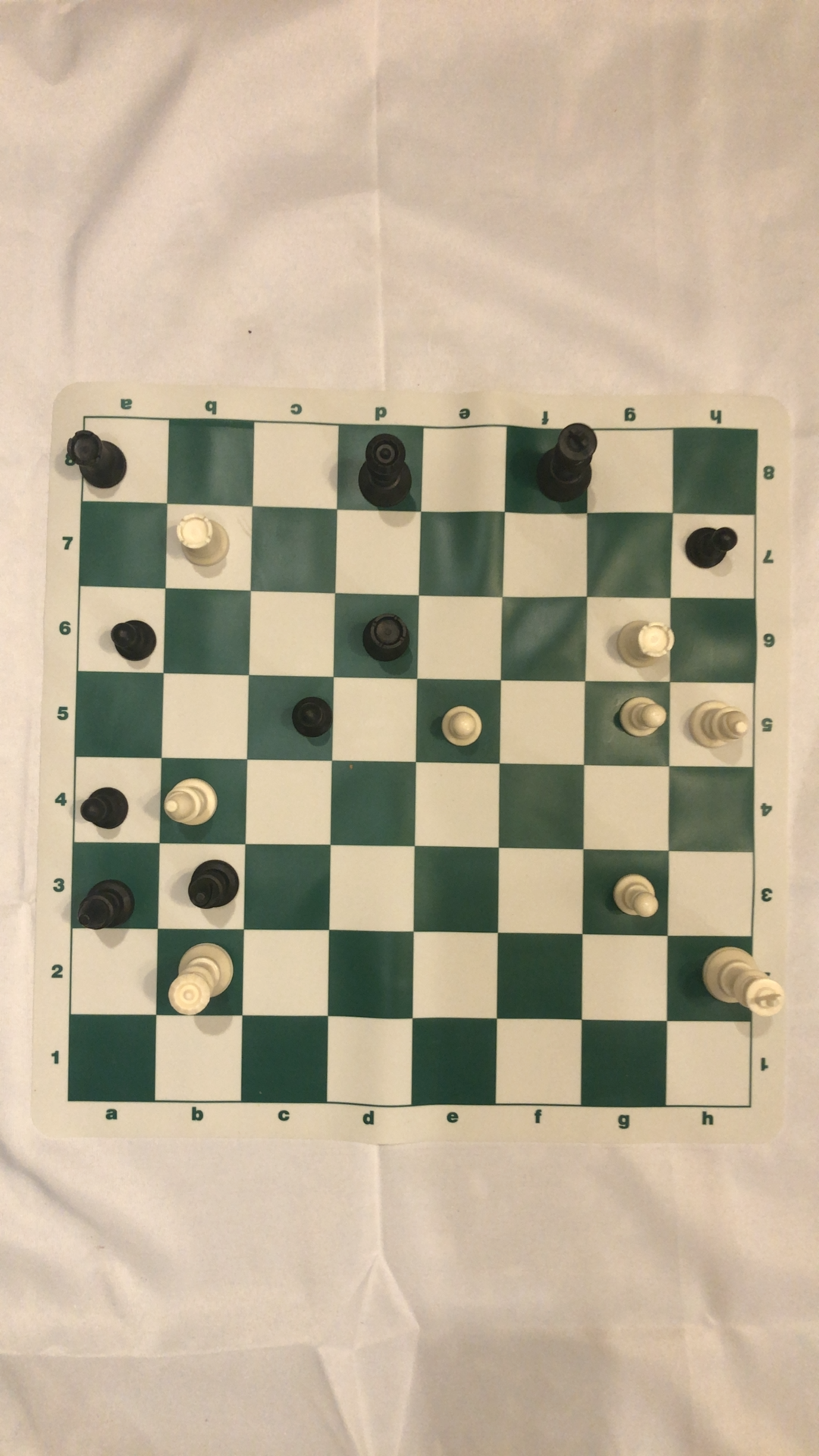} & 
      \includegraphics[width=30mm]{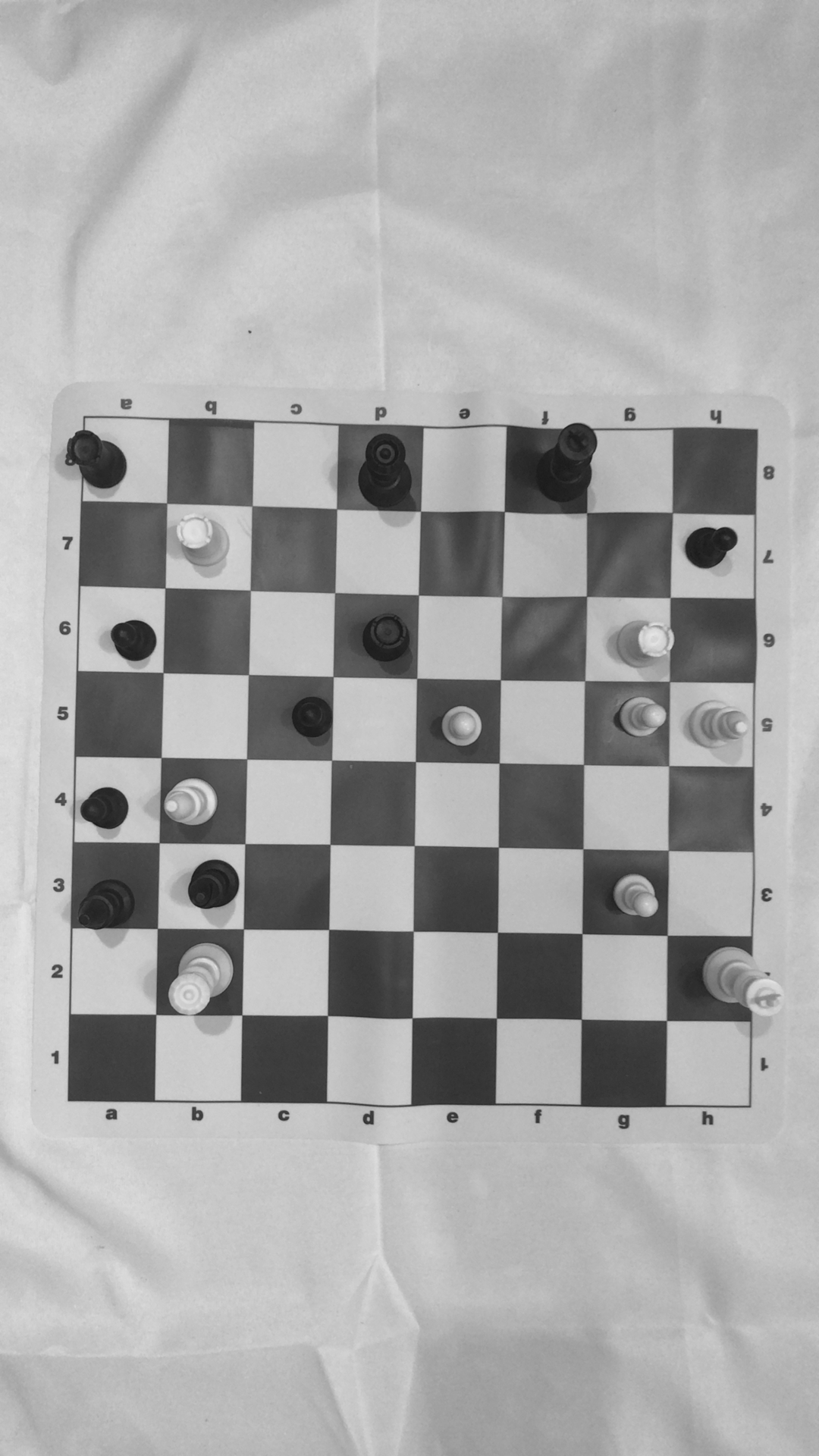} &
     \includegraphics[width=30mm]{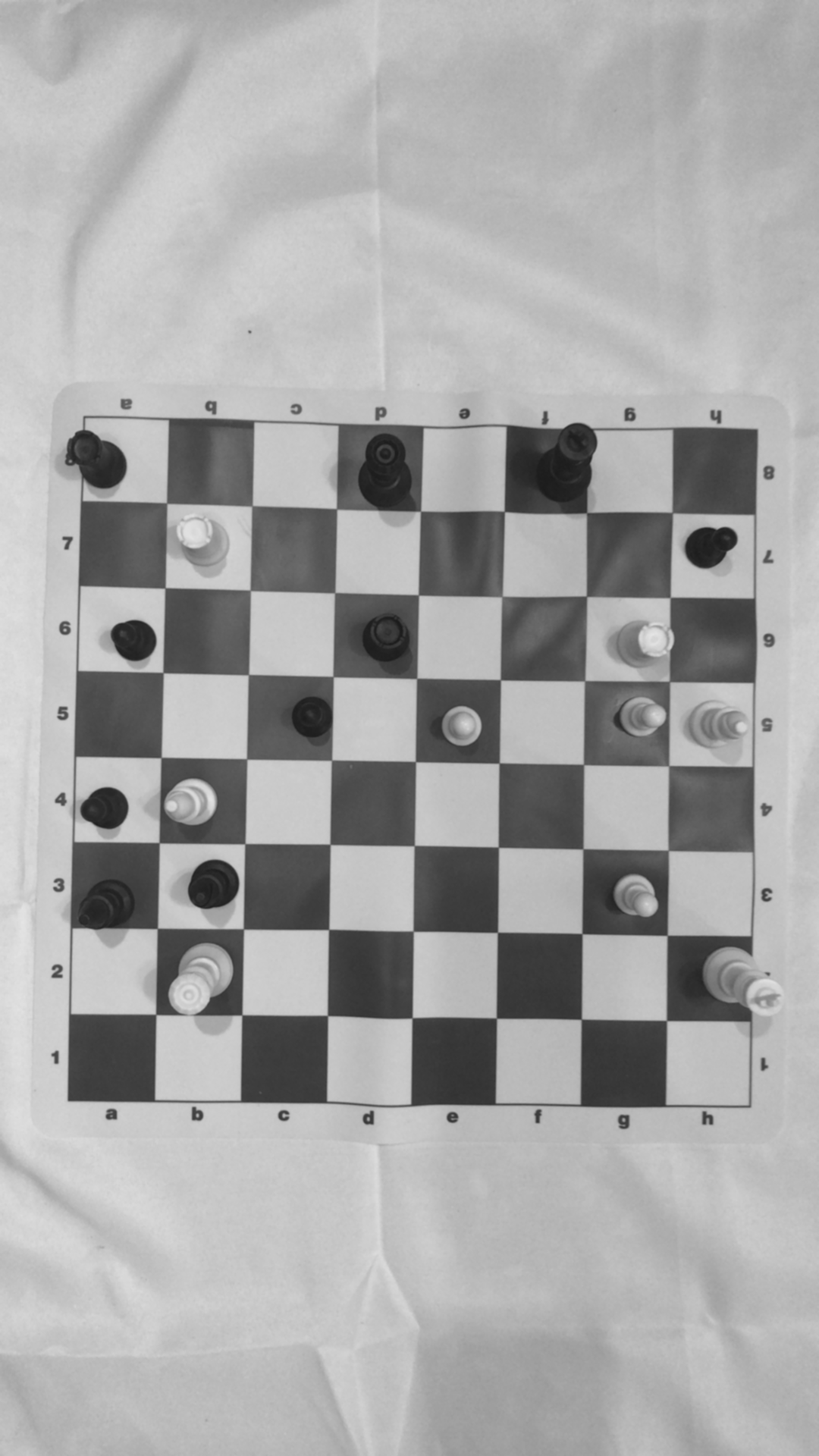} \\ \hline
         \rowcolor{LightCyan}
    Canny & Hough &  Points \\ \hline
     \includegraphics[width=30mm]{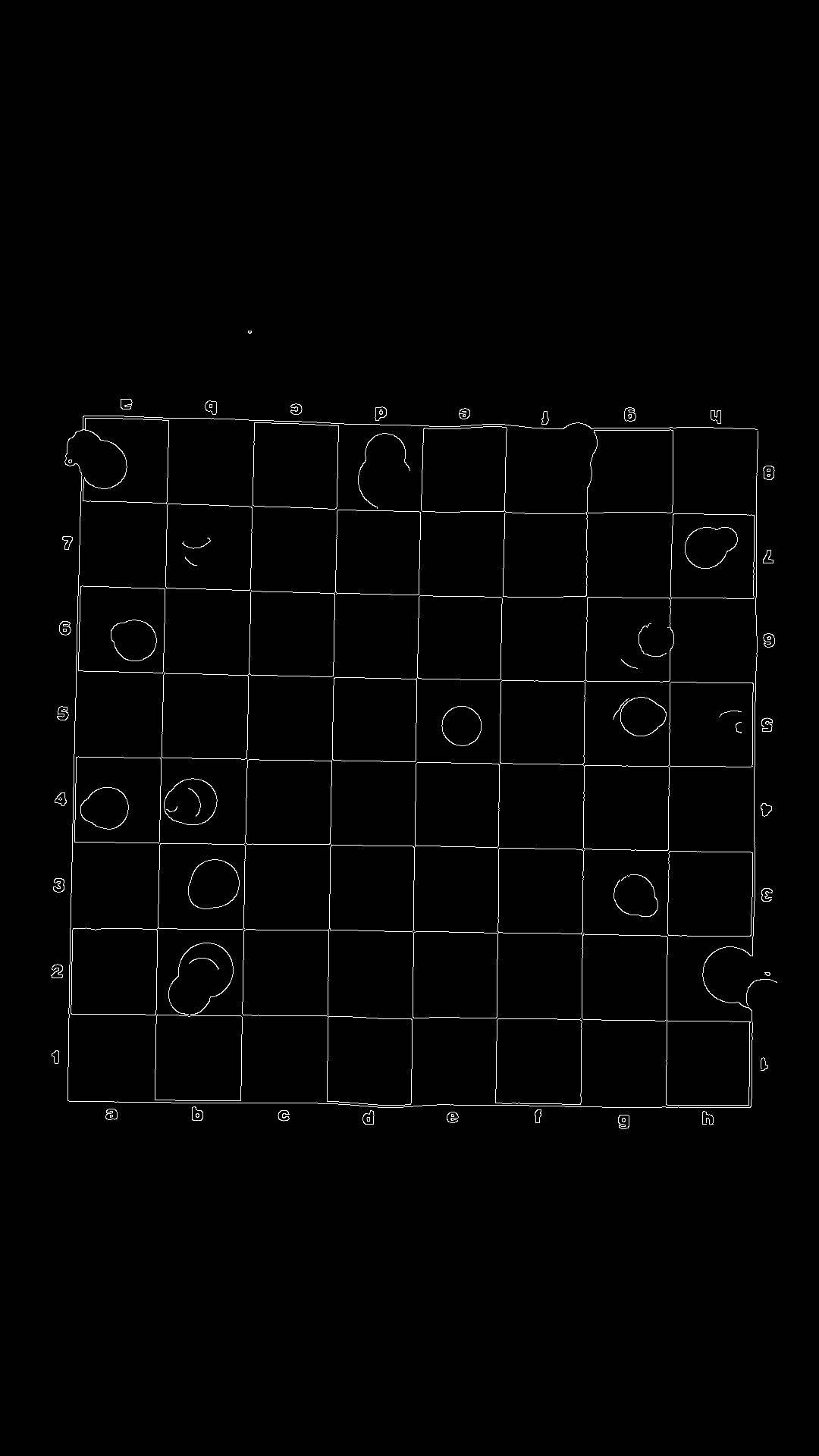} & 
      \includegraphics[width=30mm]{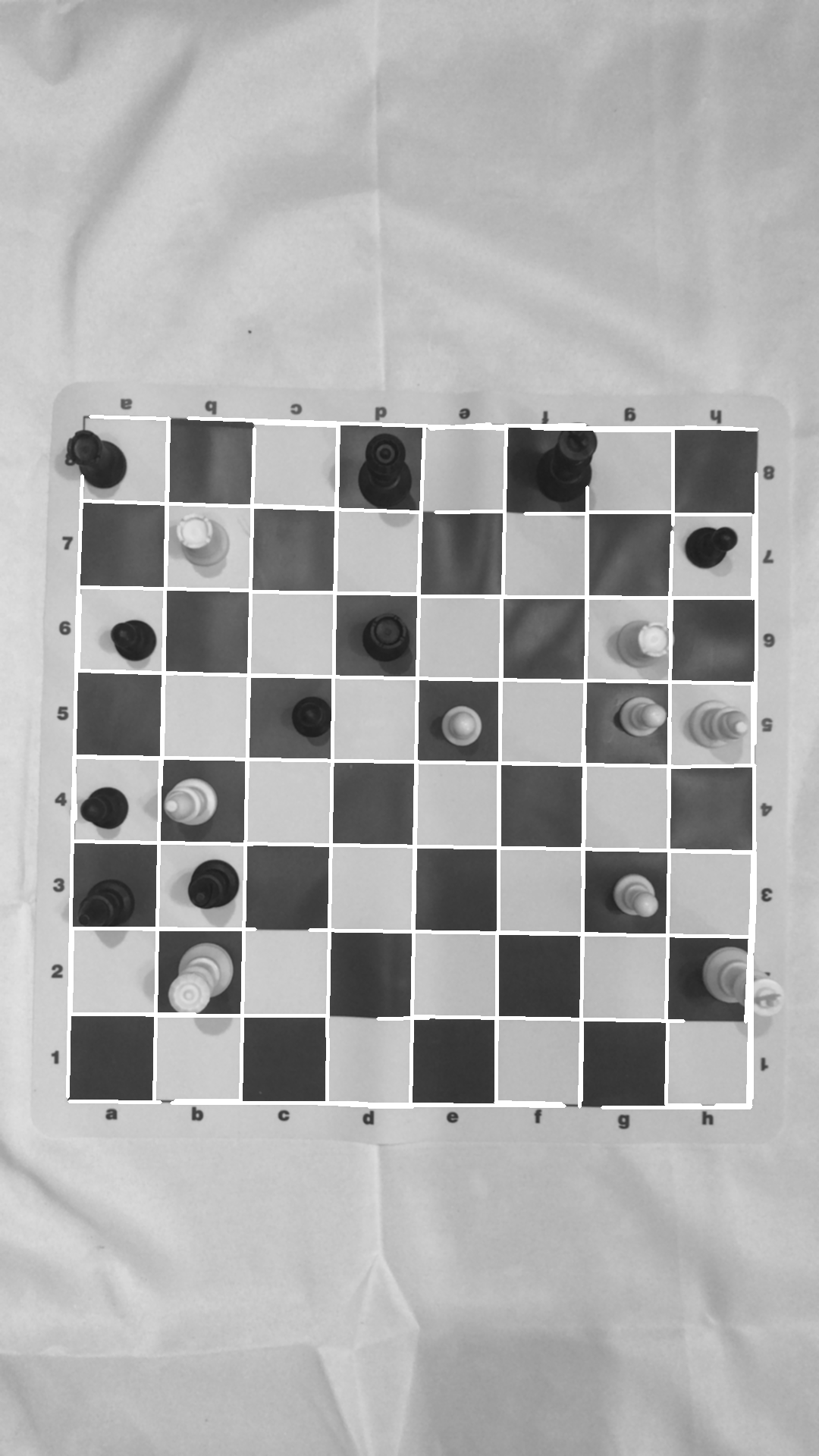} &
     \includegraphics[width=30mm]{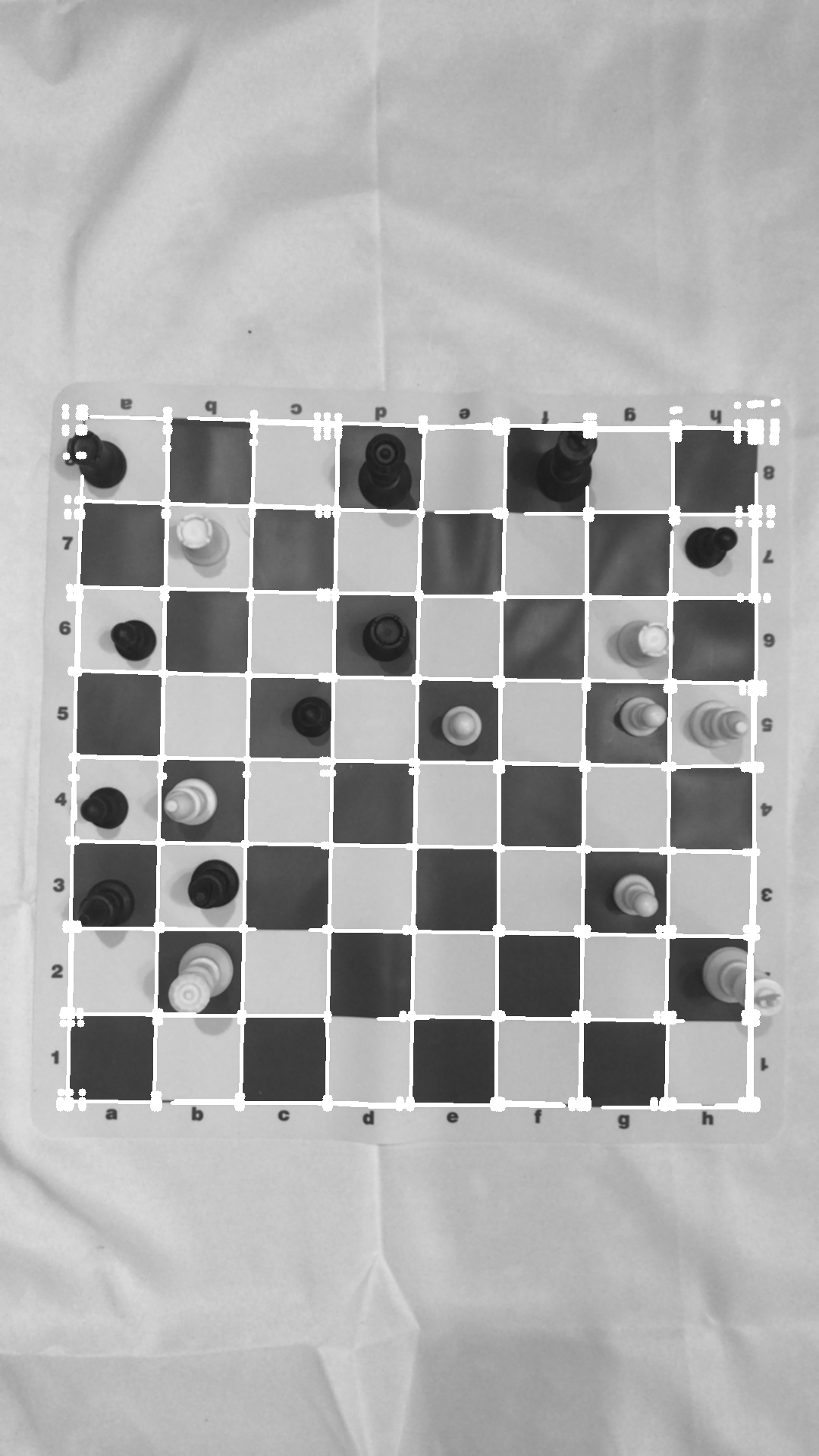} \\ \hline   
              \rowcolor{LightCyan}
        Bounded & Transformed &  Segmented \\ \hline
     \includegraphics[width=30mm]{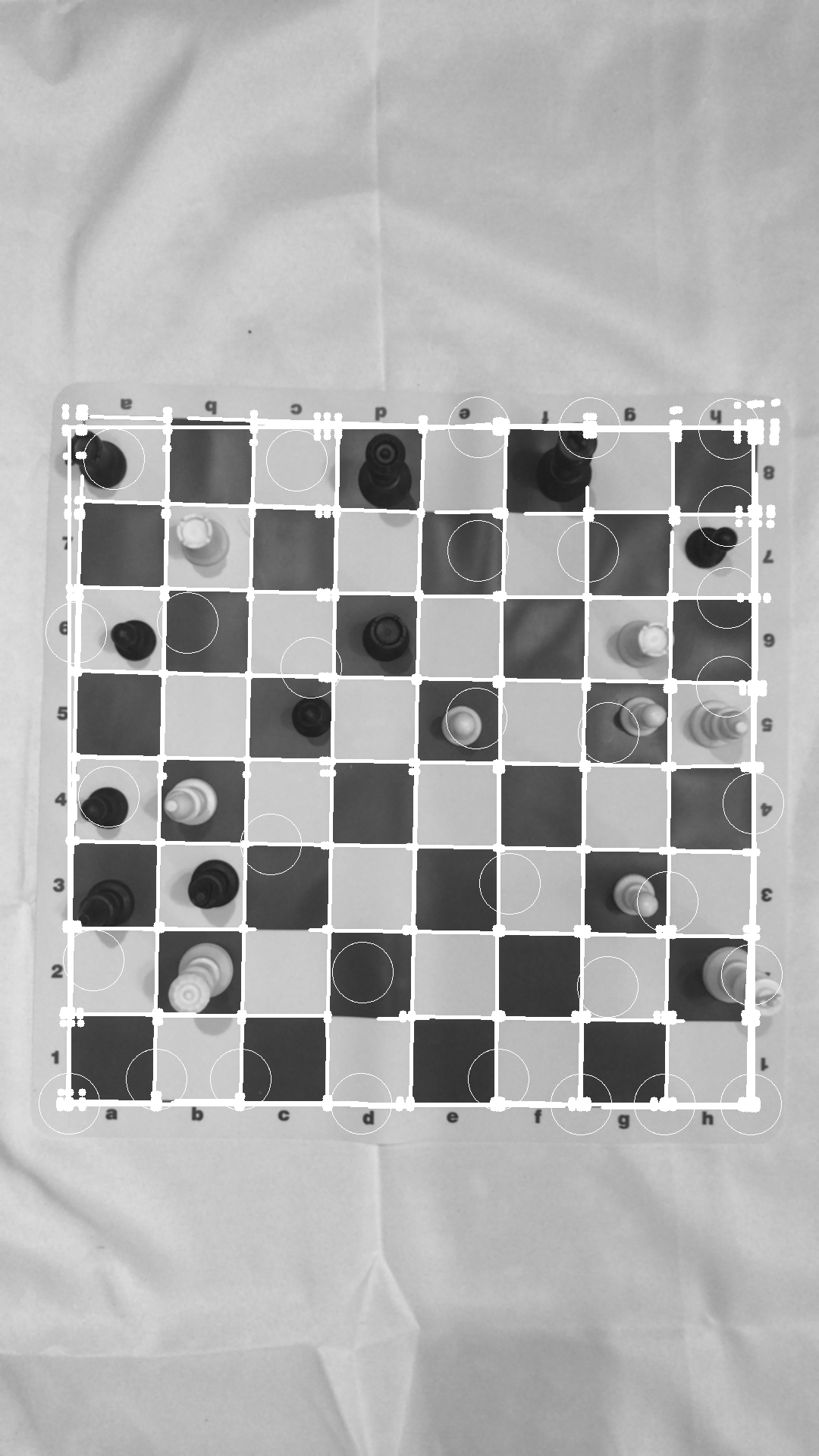} & 
      \includegraphics[width=30mm]{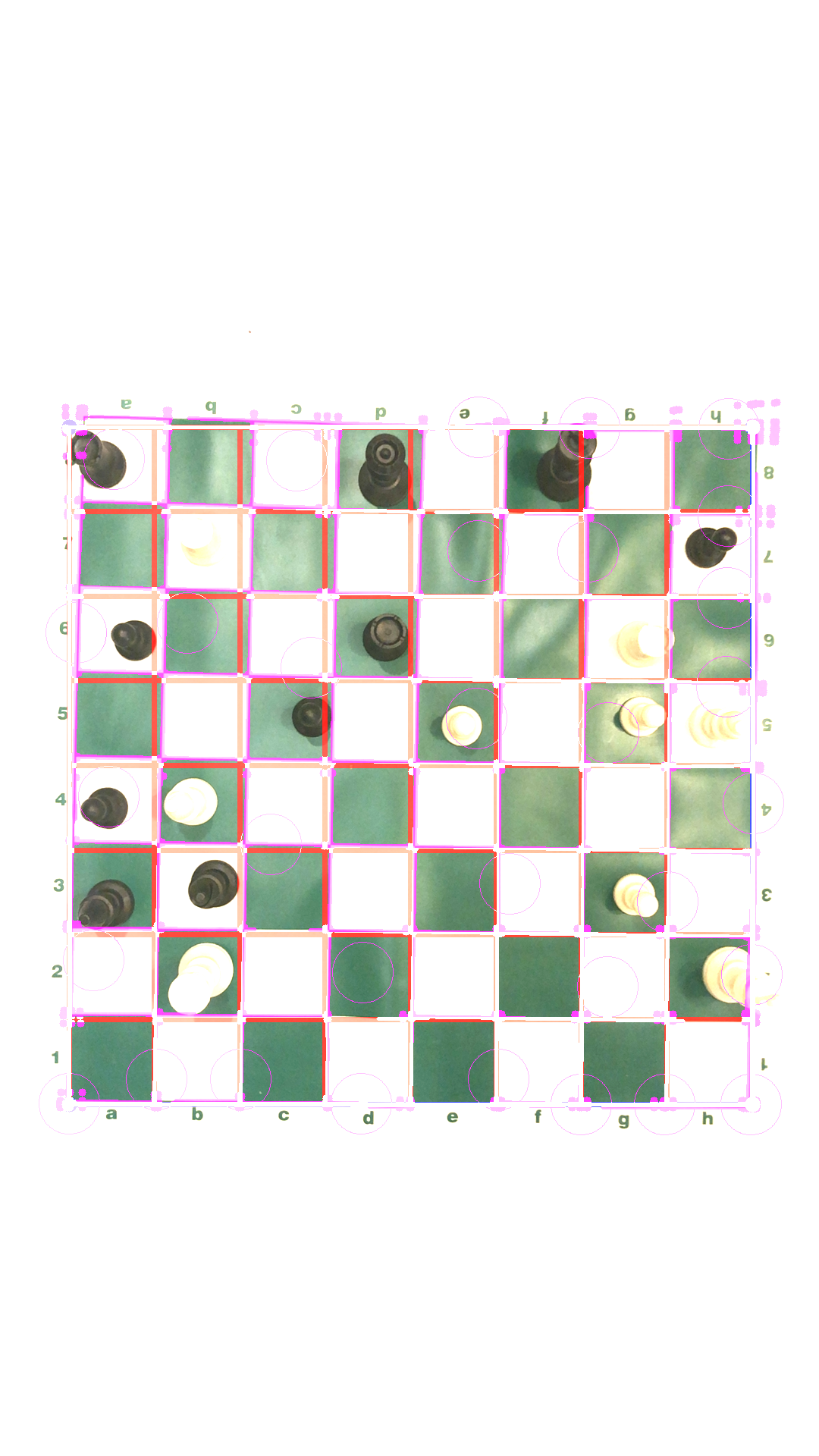} &
     \includegraphics[width=30mm]{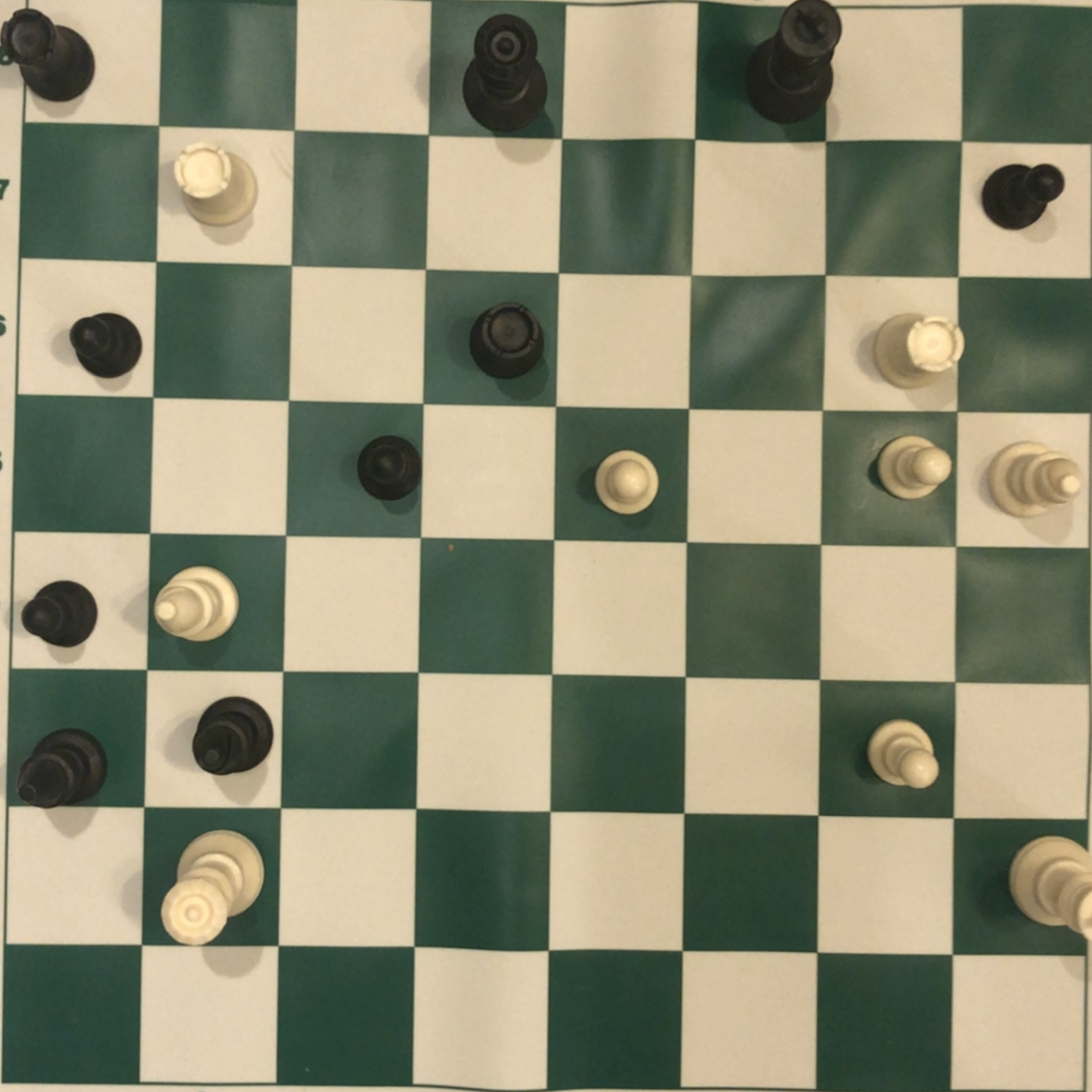} \\ \hline 
    \end{tabular}
    \label{fig:seg_pipe}
\end{figure}

\begin{itemize}
\item \textbf{Edge detection:} Smoothen the image and then use adaptive Canny edge detection similar to using gradient threshold and adaptive histogram equalization mask.
\item \textbf{Line detection:} Using HoughLines filter the segments removing small lines and merge gaps between adjacent collinear lines.
\item \textbf{Grouping:} Separate segments into groups of nearly collinear segments and merging their ends.
\item \textbf{Merging:} Analyze and merge the segments in each group utilizing the M-estimator, resulting in one normalized straight line.
\item \textbf{Filtering:} Remove non-parallel lines. In this case, the segments are bound within $10^\circ$ of horizontal and vertical
\end{itemize}

\item \textbf{Determining the Bounding box}. After the lines have been detected, the next step is finding the intersections of those lines. After the points have been determines, they are merged using K-cluster. Minimum and maximum of those intersecting points gives the box. After this, all is needed to do a projective transform to $((0,0),(227x8,227x8))$ for all the three color channels. AlexNet \cite{alexnetfig}, which is the next stage of the chess piece detection pipeline, expects the size of each image to be $227 \times 227$.. This removes any camera distortion. 


\item \textbf{Segmenting into 8x8 squares} An array of 8x8 images is created to be fed into the piece detection estimator
\end{itemize}

\subsection{Piece Detection}
\begin{figure}[h!]
    \centering
    \caption{ARChessAnalayzer model generation pipeline}
    \label{fig:model_generation}
    \includegraphics[width=\linewidth]{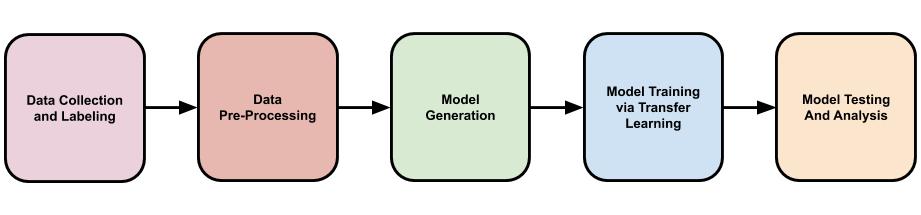}
\end{figure}

Figure \ref{fig:model_generation} summarizes the five steps in the development of the model. We will now cover each of these steps in detail.

\begin{table}[h]
    \centering

    \caption{Chess piece database}
    \begin{tiny}
    \begin{center}
    \label{table:training_label}
    \begin{tabular}{|c|c|c|c|c|c|c|c|c|c|c|c|c|c|} \hline 
    \rowcolor{LightCyan}
    \multicolumn{14}{|c|}{Label} \\ \hline
       \rowcolor{LightCyan}
    empty & br & bn & bb & bq & bk & bp & wr & wn & wb & wq & wk & wp & tot  \\ \hline \hline
    123 & 212 & 205 & 204 & 215 & 214 & 201 & 202 & 207 & 210 & 212 & 210 & 207 & 2622\\ \hline \hline
    \end{tabular}
    \end{center}
    \end{tiny}
\end{table}

\begin{figure}[h!]
\centering
\caption{Model preparation and training}
\label{fig:ModelPipeline}
\includegraphics[width=\linewidth]{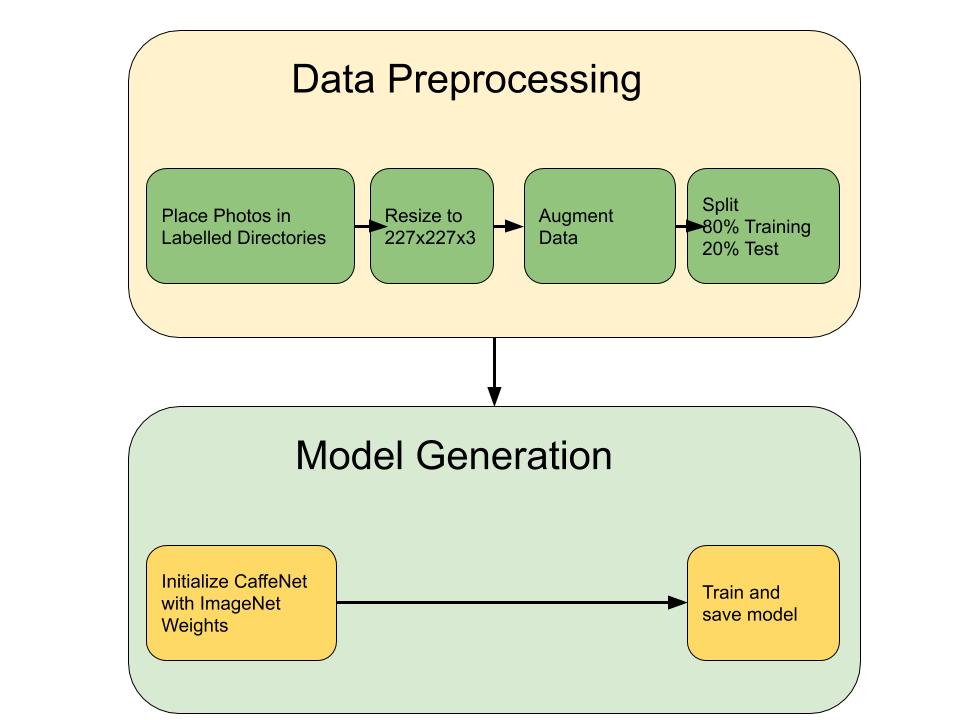} 
\end{figure}

\subsubsection{Data Collection and Labeling} 
Since there was a lack of labeled chess pieces, a database of approximately 2,600 chess pieces was manually constructed from one tournament chess set (Table \ref{table:training_label}). They were image of individual piece places on the board and manually labelled with one of 13 classes – white and black of pawn, knight, bishop, rook, queen, king, and empty (Figure \ref{tab:saliencyjpg}).

\subsubsection{Data Preprocessing}
\begin{figure} [h!]
    \centering
    \caption{Augmentation}
    \label{fig:augmentationfigures}
        \begin{tabular}{|c|c|c|c|} \hline
        \rowcolor{LightCyan}
    Orig & Cropped &  Flipped & Blurred\\ \hline \hline
      \includegraphics[width=25mm]{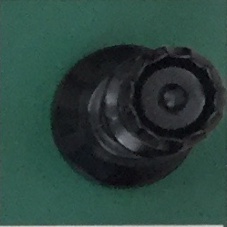} & 
      \includegraphics[width=25mm]{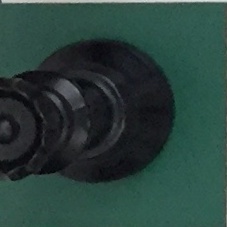} &
     \includegraphics[width=25mm]{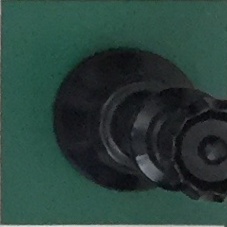} & 
      \includegraphics[width=25mm]{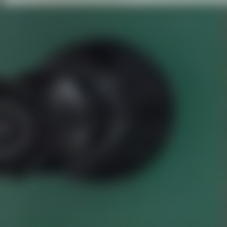} \\
        \includegraphics[width=25mm]{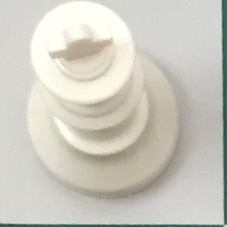} & 
      \includegraphics[width=25mm]{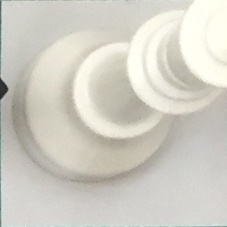} &
     \includegraphics[width=25mm]{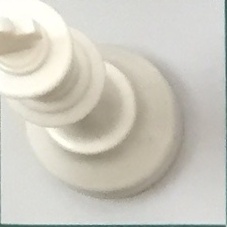} & 
      \includegraphics[width=25mm]{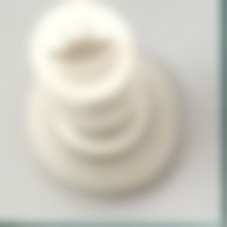} \\
      \end{tabular}
\end{figure}

Figure \ref{fig:ModelPipeline} gives a summary of the preprocessing steps. Images are resized to 227 by 227 by 3 (color channels) to fit the input of the AlexNet. To improve the performance of the CNN \cite{augm}, the data set is also augmented with transformations such as cropped, flipped and blur (Figure \ref{fig:augmentationfigures}) resulting in a total size of approx $2\times2600$. The data is then partitioned into training (80\%) and validation (20\%) sets. 

\subsubsection{Model Generation and Training}
\begin{figure}[h!]
\centering
  \caption{AlexNet \cite{alexnetfig}}
  \label{fig:AlexNet}
  \includegraphics[width=\linewidth]{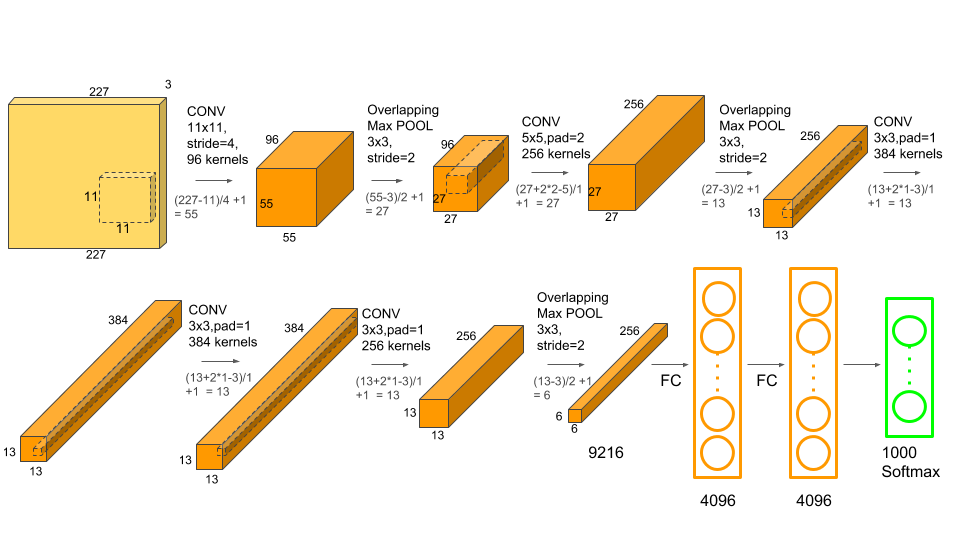}
\end{figure}

AlexNet (Figure \ref{fig:AlexNet}) is considered one of the most influential papers published, employing CNNs and GPUs to accelerate deep learning \cite{alexnetfig}. It contains eight layers; the first five were convolutional layers, some of them followed by max-pooling layers, and the last three were fully connected layers and the activation function is a SoftMax. 
While we did try other CNNs, they overfit on the data almost immediately, unlike AlexNet. This is possibly due to the combination of imbalance and small size of the data set. 

\begin{itemize}
\item \textbf{Transfer Learning}
Huge data sets with upwards of a million images are necessary to train a CNN from scratch with randomly  initialized weights. Too little data, such as our case, would cause a model to overfit.
The original AlexNet model was trained using approximately 1.3 million images (1000 object classes) from the 2012 ImageNet data set. Weights and layers from original AlexNet were used as a starting point and fine-tuned with pre-processed images with a batch size of 64. Transfer learning \cite{xfer}  leverages the previously learned low level features (such as lines, edges and curves) and requires less data to arrive at a satisfactory CNN.


\item \textbf{Model Precision}
The AlexNet model was fine tuned with 32b (FP32), but it was reduced down to 16b (FP16) with CoreML tool, during model conversion, to fit in the size of the app. The state-of-the-art hardware deep neural networks (DNNs) are moving from 32-bit computations towards 16-bit precision, due to energy efficiency and smaller associated storage. Recently \cite{8bit} showed the successful training of DNNs using 8b (FP8) while fully maintaining the accuracy on a spectrum of models and datasets. 
\end{itemize}

\subsection{Tools}
\begin{figure}[h!]
\centering
\caption{Screen shots of ARChessAnalyzer app}
\label{fig:ARChessAnalyzerapp}
\centering
\begin{tabular}{|c|c|c|} \hline 
    \rowcolor{LightCyan}
      {\small Initial screenshot} & 
      {\small Canny edges and} &
      {\small Augmented Reality} \\
      \rowcolor{LightCyan}
      & \small {Hough line segmentation} & {\small overlay}\\ \hline \hline
    \includegraphics[width=23mm]{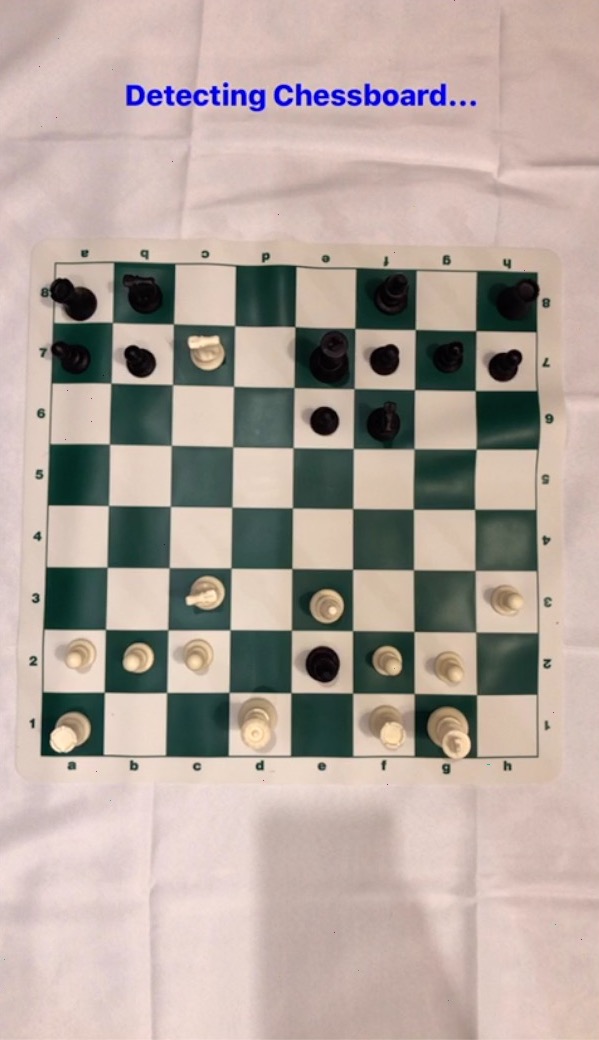} & 
      \includegraphics[width=23mm]{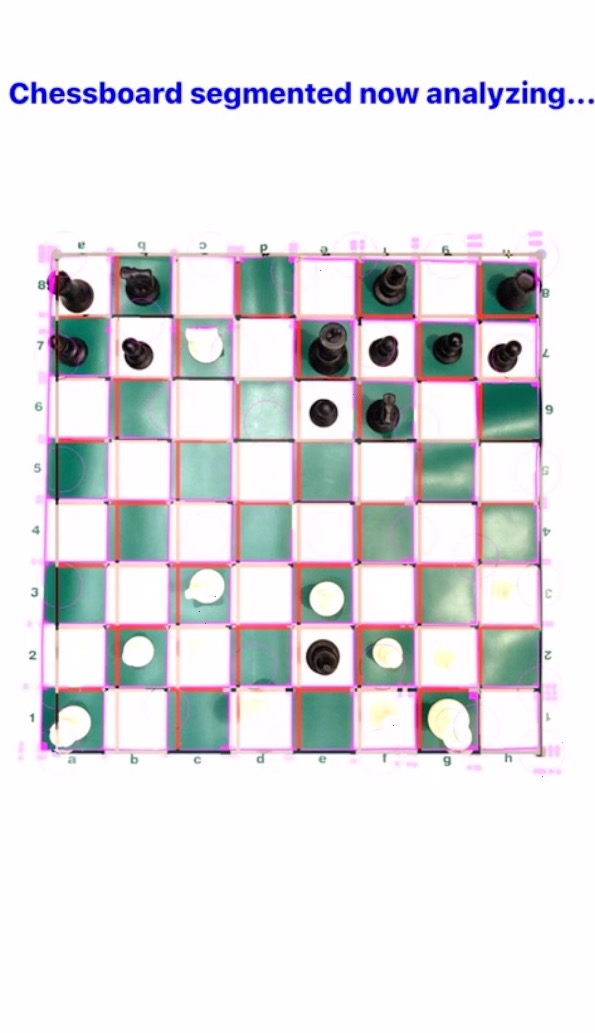} & 
      \includegraphics[width=23mm]{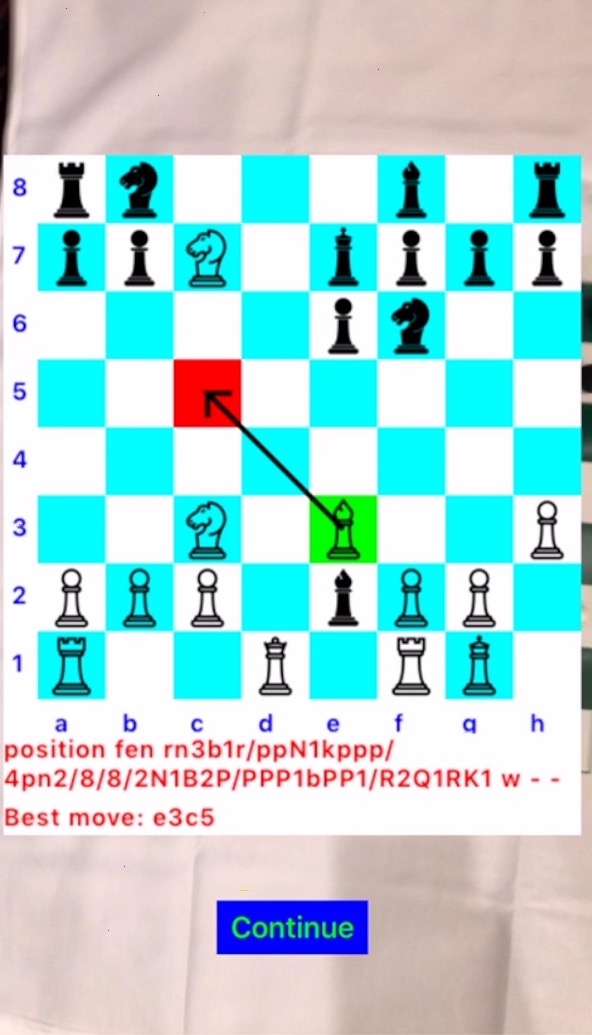} \\ \hline
          \includegraphics[width=23mm]{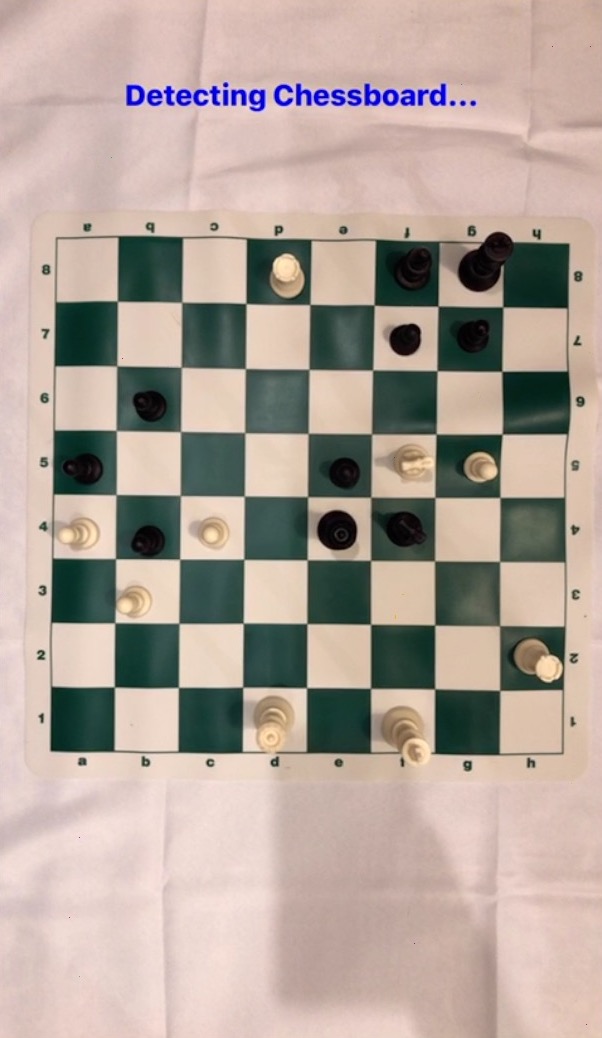} & 
      \includegraphics[width=23mm]{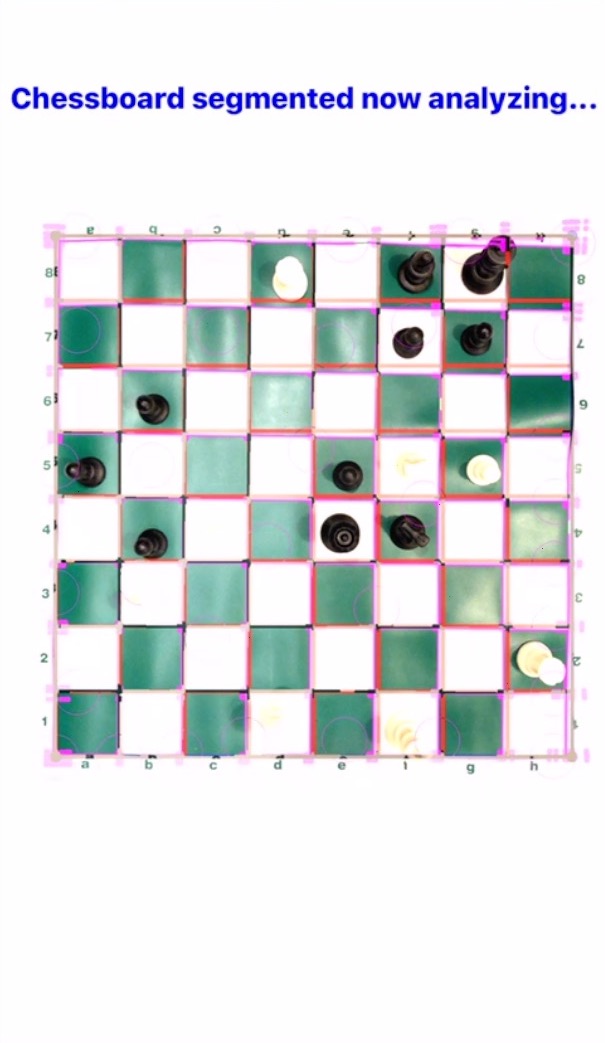} & 
      \includegraphics[width=23mm]{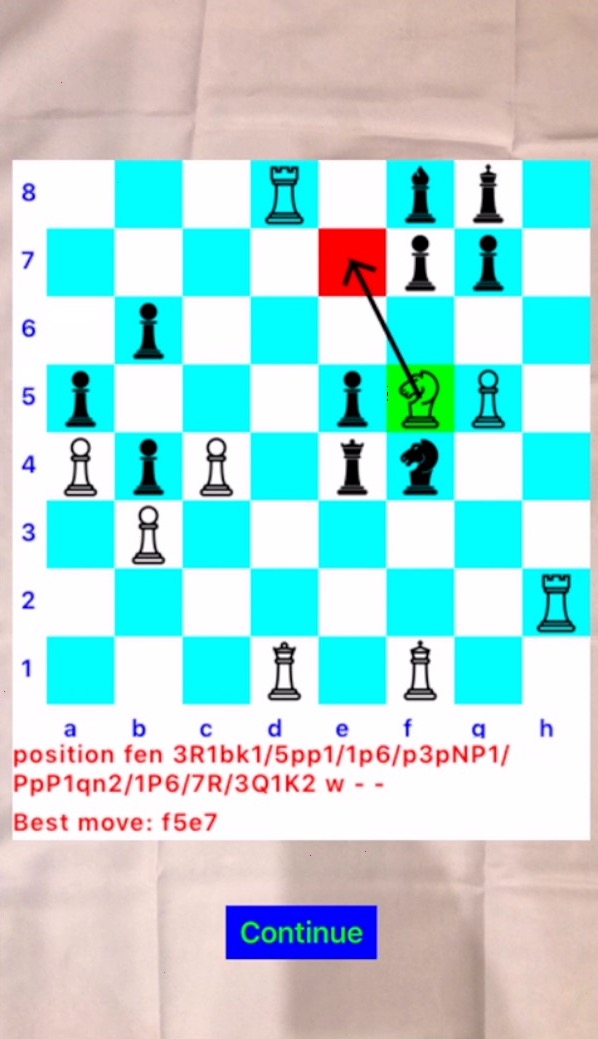} \\ \hline
          \includegraphics[width=23mm]{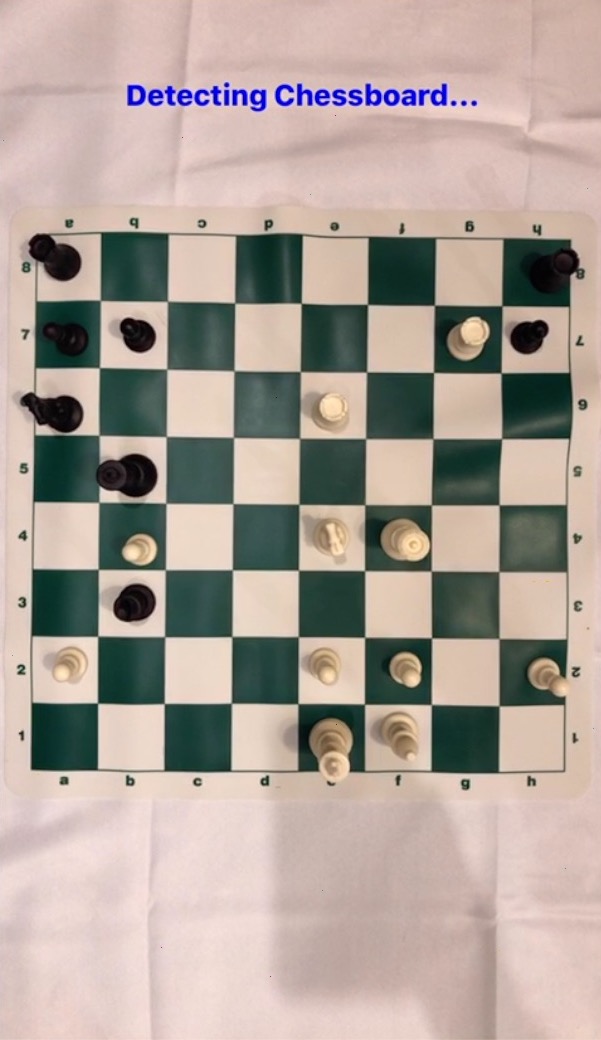} & 
      \includegraphics[width=23mm]{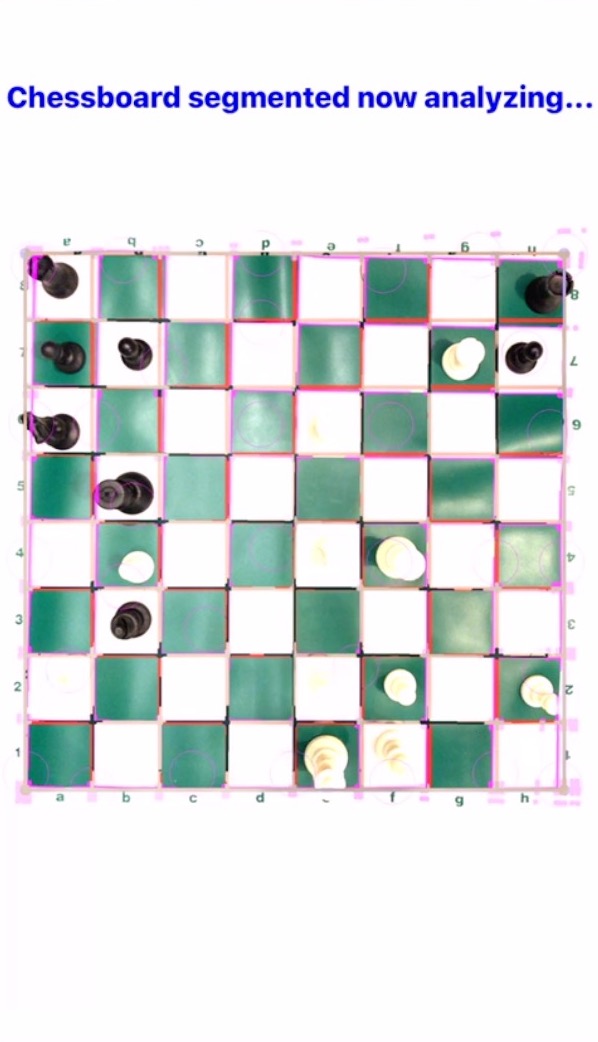} & 
      \includegraphics[width=23mm]{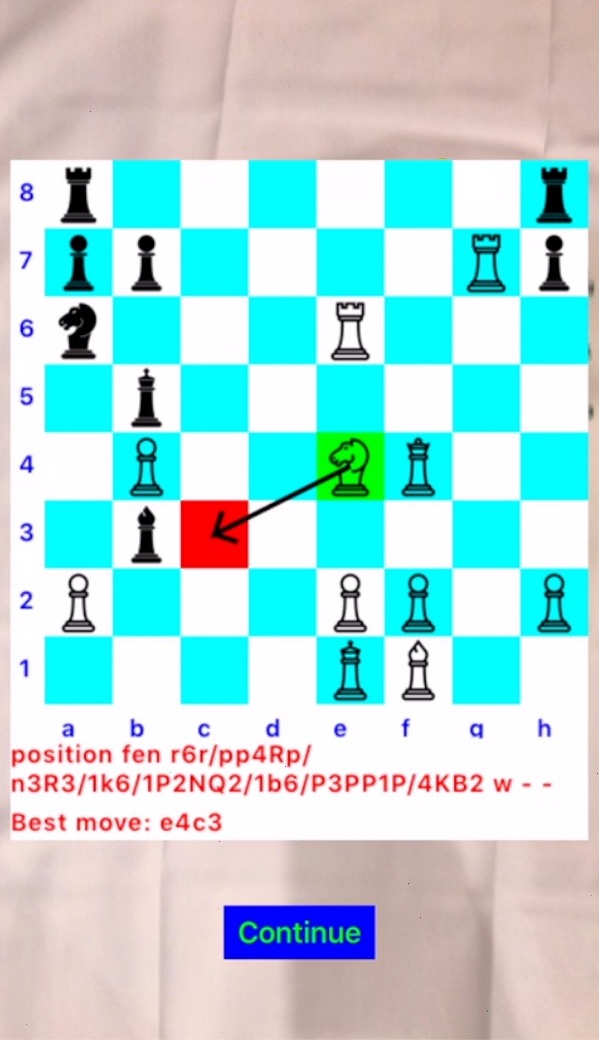} \\ \hline
          \includegraphics[width=23mm]{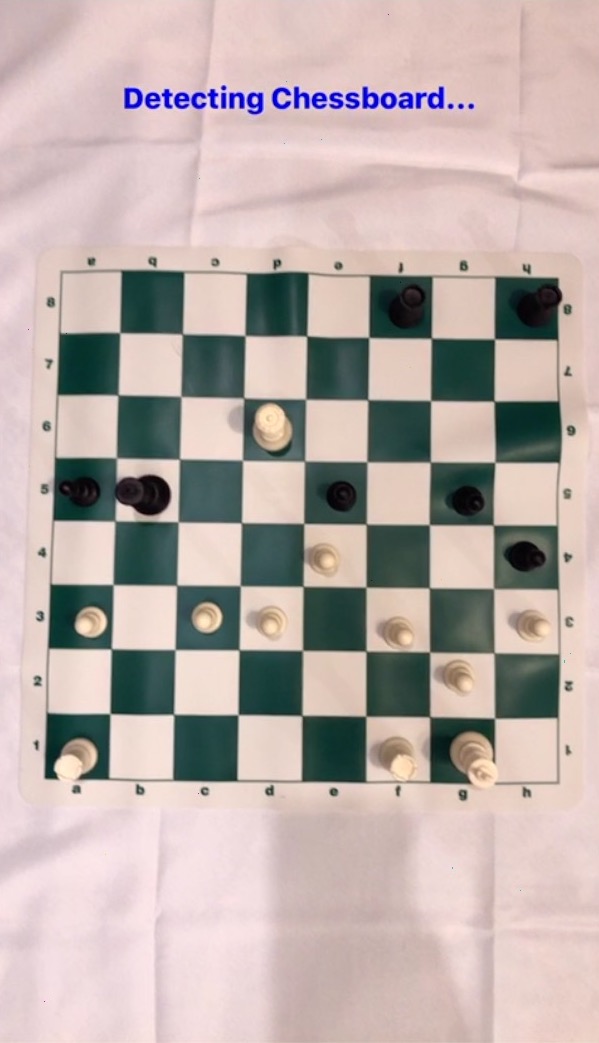} & 
      \includegraphics[width=23mm]{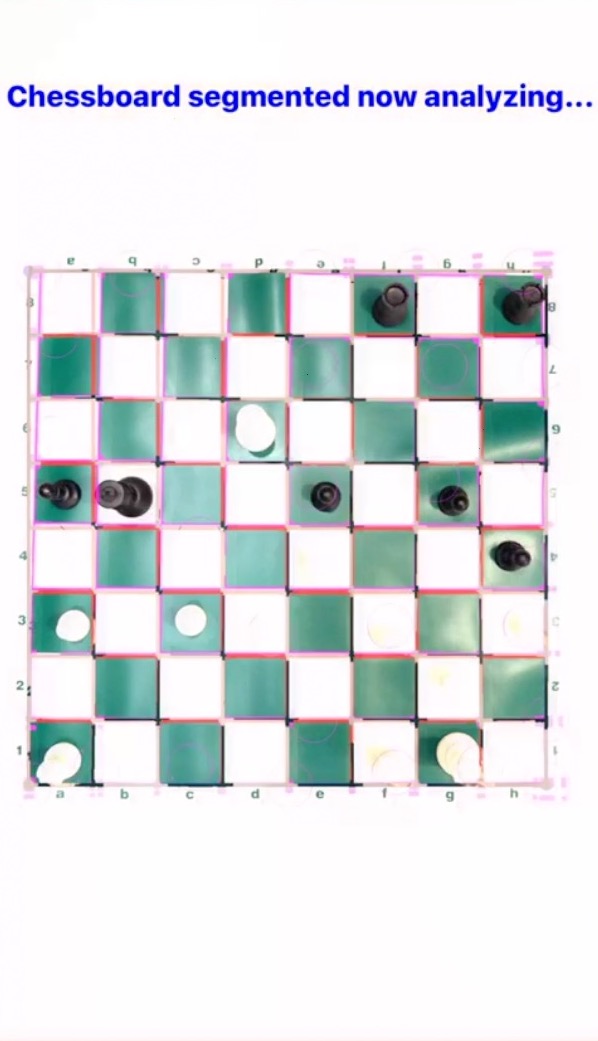} & 
      \includegraphics[width=23mm]{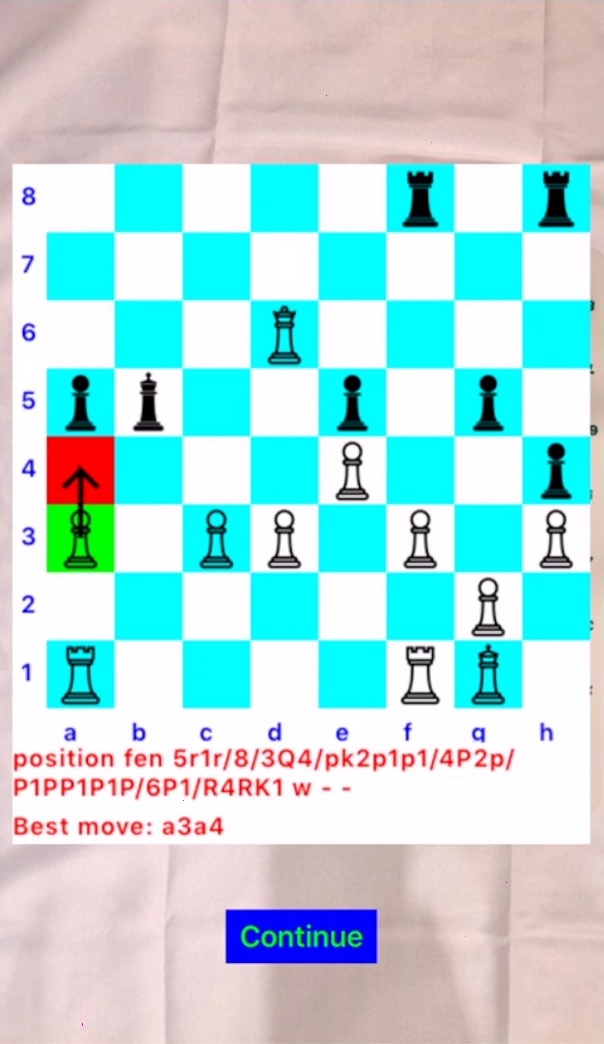} \\ \hline
 \hline
\end{tabular}
\end{figure}

Many tools went into this research. AlexNet was trained using Caffe framework with Nvidia Tesla K80 GPUs on Google Colaboratory platform using Python 3.7. The OpenCV library was used for board segmentation. A standard iOS mobile development setup was used. It was built using the Swift 5 programming language and Xcode integrated development environment with bridges to OpenCV and StockFish.
 In figure \ref{fig:ARChessAnalyzerapp} are some of the screen shots.

\newpage
\section{Experiments And Results}
A total of 720 positions were analyzed. Each board image was tagged with an expected FEN string like the following starting board position\\
rnbqkbnr/pppppppp/8/8/4P3/8/PPPP1PPP/RNBQKBNR b KQkq e3 0 1 \\
and that was compared to the following observed string. \\ rnbqkbnr/pppppppp/8/8/4$\cancelto{B}{P}$3/8/PPPP1PPP/RNBQKBNR b KQkq e3 0 1 \\
As seen above, there is a mismatch in one position - i.e. a white pawn at a center square e4 was mischaracterized as a white bishop. From this mismatch it is possible to determine the pieces which were mispredicted, the types of board positions (starting, middle, end game) and at what position in the board (edge, corners, centers) the pieces mismatched. The capture and predict routines were wrapped in a test program and results were gathered and analyzed. 

\subsection{Data Augmentation}
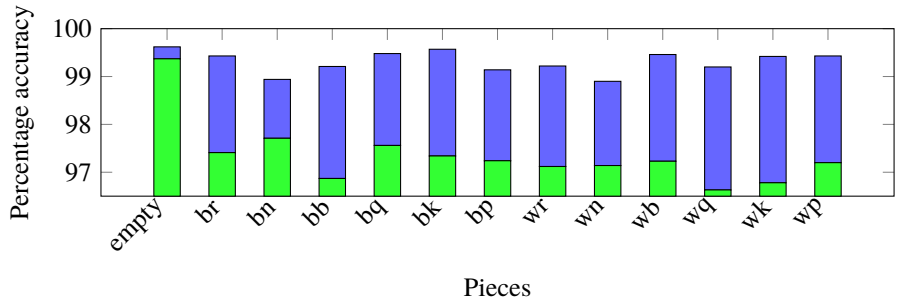
\begin{figure}[h!]
\caption{Increase in accuracy after data augmentation}
\label{fig:augment}
\centering
\begin{tikzpicture}
\begin{axis}[
        ,height=1.5in,
        ,xlabel=Pieces
        ,ylabel=Percentage accuracy
        ,ybar stacked,	ymin=96.5, ymax=100
        ,x tick label style={rotate=45,anchor=east},
        ,xtick=data
        ,xticklabels from table={\testdata}{Label}
    ]
\addplot [fill=green!80] table [y=Regular, meta=Label,x expr=\coordindex] {\testdata};   
\addplot [fill=blue!60] table [y=Augmented, meta=Label,x expr=\coordindex] {\testdata};

    \end{axis}
\end{tikzpicture}
\end{figure}

Accuracy on the piece pipeline increased approximately 1.97\% with augmentation (Fig \ref{fig:augment}). For the rest of the paper, the augmented model is used.

\pgfplotstableread{ 
Label Regular Augmented
empty 99.37 0.25
br 97.41 2.02
bn 97.71 1.23
bb 96.87 2.34
bq 97.56 1.92
bk 97.34 2.23
bp 97.24 1.90
wr 97.12 2.10
wn 97.14 1.76
wb 97.23 2.23
wq 96.63 2.57
wk 96.78 2.64
wp 97.2 2.23
    }\testdata

\subsection{Accuracy and Model Precision}
AlexNet was fine-tuned using 32b precision and later downsized to 16b (using coreml) due to size constraints of the iOS app store. The degradation in accuracy is 0.78\% (Table \ref{tab:accuracymodel}). For the rest of the paper FP16 is used. 
\begin{table}[h!]
    \centering
    \caption{Accuracy and model precision}
    \label{tab:accuracymodel}
    \begin{tabular}{|c|c|c|} \hline
    \rowcolor{LightCyan}
    AlexNet model (bit) & Size & Accuracy \\ \hline \hline
      FP32   &  221MB & 94.23\%\\ \hline
      FP16   &  106MB & 93.45\% \\ \hline
      FP8 .  &  48MB .& 93.15\% \\ \hline \hline
    \end{tabular}
\end{table}

\subsection{Saliency Maps}
\begin{figure}[h!]
    \caption{Saliency maps}
    \label{tab:saliencyjpg}
    \centering
    \begin{tabular}{|c|c|c|c|} \hline
        \rowcolor{LightCyan}
    Label & Saliency &  Label & Saliency\\ \hline \hline
    \rowcolor{LightCyan}
    \multicolumn{2}{|c|}{Black} & \multicolumn{2}{|c|}{White} \\ \hline \hline
     \rowcolor{LightCyan}
          \multicolumn{4}{|c|}{Rook}  \\ \hline \hline
      \includegraphics[width=15mm]{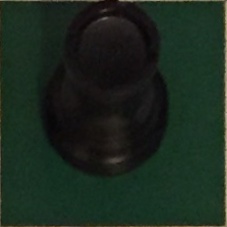} & 
      \includegraphics[width=15mm]{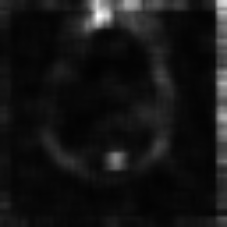} &
     \includegraphics[width=15mm]{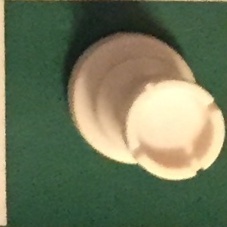} & 
      \includegraphics[width=15mm]{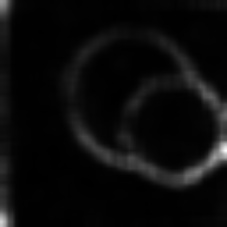} \\ \hline \hline
       \rowcolor{LightCyan}
    \multicolumn{4}{|c|}{Bishop} \\ \hline \hline
      \includegraphics[width=15mm]{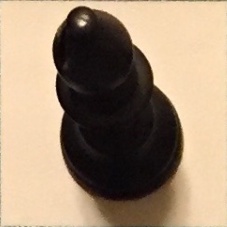} &
      \includegraphics[width=15mm]{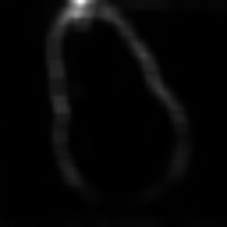} &
      \includegraphics[width=15mm]{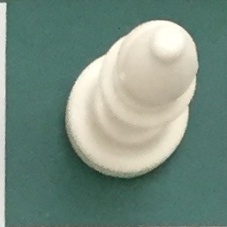} &
      \includegraphics[width=15mm]{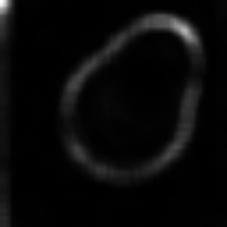} \\ \hline \hline
       \rowcolor{LightCyan}
    \multicolumn{4}{|c|}{Knight} \\ \hline \hline
      \includegraphics[width=15mm]{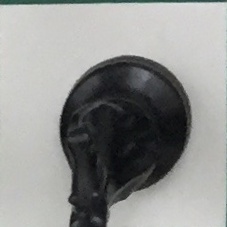} & 
      \includegraphics[width=15mm]{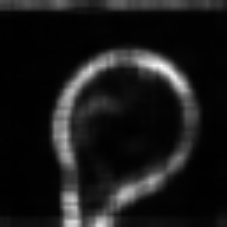} & 
     \includegraphics[width=15mm]{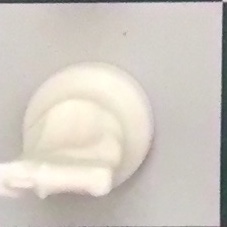} &
      \includegraphics[width=15mm]{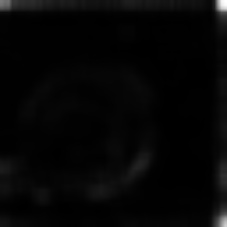} \\ \hline \hline
       \rowcolor{LightCyan}
     \multicolumn{4}{|c|}{King} \\ \hline \hline
      \includegraphics[width=15mm]{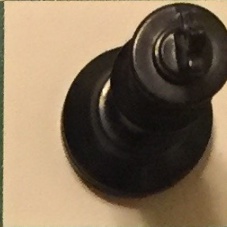} &
      \includegraphics[width=15mm]{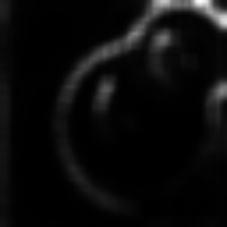} & 
     \includegraphics[width=15mm]{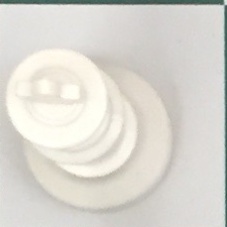} &
      \includegraphics[width=15mm]{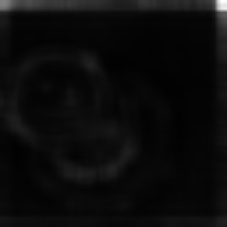} \\ \hline \hline
       \rowcolor{LightCyan}
      \multicolumn{4}{|c|}{Queen} \\ \hline \hline
      \includegraphics[width=15mm]{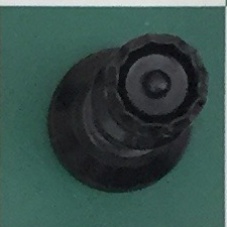} &
      \includegraphics[width=15mm]{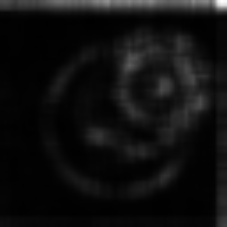} &
     \includegraphics[width=15mm]{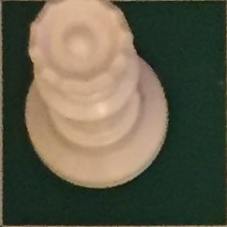} &
      \includegraphics[width=15mm]{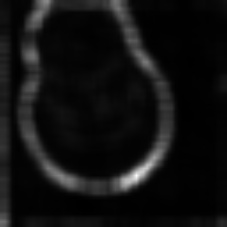} \\ \hline \hline
       \rowcolor{LightCyan}
      \multicolumn{4}{|c|}{Pawn} \\ \hline \hline
      \includegraphics[width=15mm]{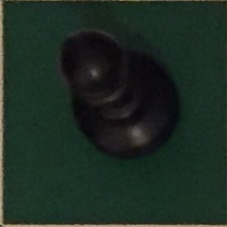} &
      \includegraphics[width=15mm]{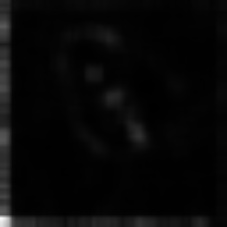} & 
      \includegraphics[width=15mm]{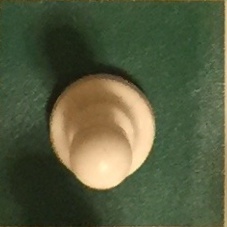} &
      \includegraphics[width=15mm]{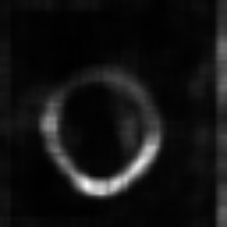} \\ \hline \hline
    \end{tabular}
\end{figure}

Saliency maps help visualize the inner workings of CNNs via a heat map highlighting the image features that the model is focused on \cite{saliency}. Saliency maps were generated on a few images to confirm that the classifiers are identifying the correct regions of interest for a particular labels (Figure \ref{tab:saliencyjpg}). 

\subsection{Classification Metrics}
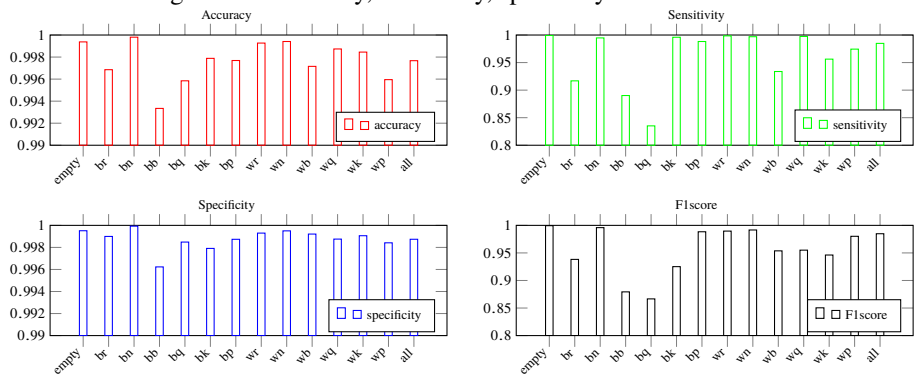
\begin{figure}[h!]
    \centering
    \begin{tiny}
    \caption{Accuracy, sensitivity, specificity and F1score}
    \label{fig:assf}
    \begin{tabular}{cc}
    Accuracy & Sensitivity \\
\begin{tikzpicture}
    \begin{axis}[
        bar width=0.1cm,
            ybar,
            height=1.2in,
            width=2.7in,
            symbolic x coords={empty,br,bn,bb,bq,bk,bp,wr,wn,wb,wq,wk,wp,all},
            xtick=data,
            legend style={anchor=south west, legend pos=south east},
            x tick label style={rotate=45,anchor=east},
            yticklabel style={/pgf/number format/precision=3},
            ymin=0.99,ymax=1,
        ]
        \addplot [color=red,solid] table[x=Label,y=accuracy]{\myconfusiondata};
        \legend{accuracy}
    \end{axis}
\end{tikzpicture} & 
\begin{tikzpicture}
    \begin{axis}[
        bar width=0.1cm,
            ybar,
            height=1.2in,
            width=2.7in,
            symbolic x coords={empty,br,bn,bb,bq,bk,bp,wr,wn,wb,wq,wk,wp,all},
            xtick=data,
            legend style={anchor=south west, legend pos=south east},
            x tick label style={rotate=45,anchor=east},
            ymin=0.8,ymax=1,
        ]
        \addplot [color=green,solid] table[x=Label,y=sensitivity]{\myconfusiondata};
        \legend{sensitivity}
    \end{axis}
\end{tikzpicture} \\
Specificity & F1score \\
\begin{tikzpicture}
    \begin{axis}[
        bar width=0.1cm,
            ybar,
            height=1.2in,
            width=2.7in,
            symbolic x coords={empty,br,bn,bb,bq,bk,bp,wr,wn,wb,wq,wk,wp,all},
            xtick=data,
            legend style={anchor=south west, legend pos=south east},
            x tick label style={rotate=45,anchor=east},
            ymin=0.99,ymax=1,
            yticklabel style={/pgf/number format/precision=3},
        ]

        \addplot [color=blue,solid] table[x=Label,y=specificity]{\myconfusiondata};

        \legend{specificity}
    \end{axis}
\end{tikzpicture} &
\begin{tikzpicture}
    \begin{axis}[
        bar width=0.1cm,
            ybar,
            height=1.2in,
            width=2.7in,
            symbolic x coords={empty,br,bn,bb,bq,bk,bp,wr,wn,wb,wq,wk,wp,all},
            xtick=data,
            legend style={anchor=south west, legend pos=south east},
            x tick label style={rotate=45,anchor=east},
            ymin=0.8,ymax=1,
        ]

        \addplot [color=black,solid] table[x=Label,y=F1score]{\myconfusiondata};
        \legend{F1score}
    \end{axis}
\end{tikzpicture}

    \end{tabular}
    \end{tiny}
\end{figure}

Generalization properties for multiclass classification derived from the confusion matrix (Table \ref{tab:Confusion}) of a classifier are used as a measure of its quality. The following performance metrics are evaluated - accuracy (Eq 2), sensitivity (Eq 3), specificity (Eq 4), and F1 Score (Eq 5). Figure \ref{fig:assf} displays all of the scores. 

\begin{small}

\begin{equation}
\text{accuracy} = \frac{\text{all correct}}{\text{all samples}}
= \frac{\text{true pos+true neg}}{\text{true pos+true neg+false pos+false neg}}
\end{equation}

\begin{equation}
\text{sensitivity} = \frac{\text{true pos}}{\text{all pos}} = \frac{\text{true pos}}{\text{true pos}+\text{false neg}}
\end{equation}

\begin{equation}
\text{specificity} = \frac{\text{true neg}}{\text{all neg}} = \frac{\text{true neg}}{\text{true neg}+\text{false pos}}
\end{equation}

\begin{equation}
\text{F1score} = 2 \times \frac{\text{sensitivity} \times \text{specificity}}
{\text{sensitivity} + \text{specificity}}
\end{equation}
\end{small}

\begin{table}[h]
\centering
\caption{Confusion matrix between actual and predicted pieces}
\begin{tiny}
\label{tab:Confusion}
\begin{center}
\begin{tabular}{|c|c|c|c|c|c|c|c|c|c|c|c|c|c|} \hline 
\rowcolor{LightCyan}
\multicolumn{14}{|c|}{Predicted cells} \\ \hline 
\rowcolor{LightCyan}
& empty&br& bn & bb & bq & bk & bp & wr & wn & wb & wq & wk & wp\\ \hline \hline
empty &	31454&	5	&4	&5	&0	&1	&4	&4	&0	&0	&0	&0	&0 \\ \hline
br&	0&	1345&	0&	50&	16&	21&	35&	0&	0&	0&	0&	0&	0\\ \hline
bn&	0&	0&	1330&	1&	4&	2&	0&	0&	0&	0&	0&	0&	0\\ \hline
bb&	4&	50&	0&	1370&	54&	37&	24&	0&	0&	0&	0&	0&	0\\ \hline
bq&	0&	0&	0&	100&	760&	50&	0&	0&	0&	0&	0&	0&	0\\ \hline
bk&	3&	0&	0&	0&	0&	735&	0&	0&	0&	0&	0&	0&	0\\ \hline
bp&	0&	0&	0&	51&	10&	5&	5540&	0&	0&	0&	0&	0&	0\\ \hline
wr&	3&	0&	0&	0&	0&	0&	0&	1955&	0&	0&	0&	0&	0\\ \hline
wn&	0&	0&	0&	0&	0&	0&	1&	0&	1955&	0&	0&	5&	0\\ \hline
wb&	0&	0&	0&	0&	0&	0&	0&	0&	0&	1650&	35&	37&	45\\ \hline
wq&	2&	0&	0&	0&	0&	0&	0&	0&	0&	0&	754&	0&	0\\ \hline
wk&	0&	0&	0&	0&	0&	0&	0&	0&	0&	0&	0&	765&	35\\ \hline
wp&	0&	0&	0&	0&	0&	0&	0&	34&	27&	43&	34&	10&	5634\\ \hline \hline
\end{tabular}
\end{center}
\end{tiny}
\end{table}

\pgfplotstableread[row sep=\\,col sep=&]{
    Label & accuracy & sensitivity & specificity & F1score \\
empty &    0.9993764809	& 0.9992693077	& 0.999513303	& 0.9994439413 \\
br & 0.9968547312	& 0.9168370825	& 0.9989964969	& 0.9382629927 \\
bn & 0.999803953	& 0.9947643979	& 0.99992697	& 0.9958816923 \\
bb & 0.9933420689	& 0.890188434	& 0.9962319104	& 0.8793324775 \\
bq & 0.9958460555	& 0.8351648352	& 0.9984843564	& 0.8665906499 \\
bk & 0.9978832026	& 0.9959349593	& 0.9979091188	& 0.9251101322 \\
bp & 0.9976879846	& 0.9882268998	& 0.9987357275	& 0.9884032114 \\
wr & 0.9992696699	& 0.9984678243	& 0.9992986471	& 0.9896228803 \\
wn & 0.9994120896	& 0.9969403366	& 0.9995015691	& 0.991630738 \\
wb & 0.99715596	& 0.9337860781	& 0.9992108789	& 0.9537572254 \\
wq & 0.9987359576	& 0.9973544974	& 0.9987548048	& 0.9550348322 \\
wk & 0.998451544	& 0.95625	& 0.9990611176	& 0.9461966605 \\
wp & 0.9959521358	& 0.9744033207	& 0.9984172206	& 0.9801670146 \\
all & 0.9976716062	& 0.9848301187	& 0.9987390314	& 0.9848301187 \\
    }\myconfusiondata


\subsection{Accuracy of Overall Pipeline}
The overall accuracy of the  position prediction string is 93.45\%. A detailed analysis of the mismatched FEN strings was done considering the types of pieces on the board, the distribution of the pieces on the board and the population of the board,.  There is high rate of mis-classification among certain types of pieces especially between queen, bishop, pawn and king \ref{tab:Confusion}. Secondly, boards with center distribution had a higher accuracy than those at the edges or corners (Table \ref{tab:accuracy_pos}). This is because of the bound clipping and projective transform and less training images of pieces from the sides. The last source of error is piece population (Opening, Midgame, Endgame) (Table \ref{tab:accuracy_exp_obs}) This is possibly because as the game proceeds from opening to end, there are less mis-classification possibilities. 
More training images especially among pieces with high rate of mis-classification and also not just using the top view but also from the sides, a slight board expansion after segmentation to incorporate the edge ring could help capture all the features, and a two-step window algorithm processing the edge and corner pieces separately would possibly alleviate these issues. We will this address this in our future work.

\begin{table}[h!]
\caption{Accuracy of piece prediction based on position}
\label{tab:accuracy_pos}
\centering
\begin{tabular}{|c|c|} \hline
\rowcolor{LightCyan}
Position & Accuracy \\ \hline \hline
Corner & 91.45\% \\
Edge &  92.87\% \\
Center & 96.73\% \\ \hline \hline
\rowcolor{LightCyan}
Total & 93.45\% \\ \hline \hline
\end{tabular}

\begin{tiny}
\begin{center}
\begin{tablenotes}

\item[1] Corner: More than 50\% of the four corners are occupied
\item[2] Edge: If not Corner and more than 50\% of the 28 edges are occupied
\item[3] Center: If not Corner or Edge

  \end{tablenotes}
    \end{center}
  \end{tiny}

\end{table}


\begin{table}[h!]
\begin{tiny}
    \centering
    \caption{Model of positions calculated expected and actual accuracy}
    \label{tab:accuracy_exp_obs}
    \begin{tabular}{|c|c|c|c|c|c|c|c|c|c|c|c|c|c|c|c|}\hline
    \rowcolor{LightCyan}
    Pos & empty & br & bn & bb & bq & bk & bp & wr & wn & wb & wq & wk & wp & Exp \%	& Act \%\\ \hline \hline
    \rowcolor{LightYellow}
	Open&32	&2	&2	&2	&1	&1	&8	&2	&2	&2	&1	&1	&8	&89.72 & 91.02\\
	\rowcolor{LightYellow}
	-ing&32	&2	&2	&2	&1	&1	&6	&2	&2	&2	&1	&1	&6	&90.88 & \\
		\rowcolor{LightYellow}
	&39	&2	&2	&2	&1	&1	&5	&1	&2	&2	&1	&1	&5	&91.12 &\\
		\rowcolor{LightYellow}
	&42	&1	&2	&2	&1	&1	&4	&2	&2	&2	&1	&1	&3	&92.13 &\\
		\rowcolor{LightYellow}
	&43	&0	&2	&2	&1	&1	&4	&2	&2	&2	&1	&1	&3	&92.36 &\\ \hline
    \rowcolor{LightBlue}
	Mid-&45	&1	&1	&2	&1	&1	&3	&2	&1	&2	&1	&1	&3	&92.24& 93.47\\
	\rowcolor{LightBlue}
	game&48	&1	&1	&1	&1	&1	&3	&1	&1	&1	&1	&1	&3	&93.02 &\\
	\rowcolor{LightBlue}
	&49	&0	&1	&1	&1	&1	&3	&2	&1	&1	&1	&1	&2	&93.57 &\\
	\rowcolor{LightBlue}
	&49	&1	&1	&1	&1	&1	&2	&2	&1	&1	&1	&1	&2	&93.49 &\\
	\rowcolor{LightBlue}
	&47	&1	&1	&1	&1	&1	&3	&2	&1	&1	&1	&1	&3	&93.01 & \\ \hline
	\rowcolor{LightPurple}
	End-&56	&1	&0	&0	&1	&1	&1	&1	&0	&0	&1	&1	&1	&94.71 & 95.24\\
		\rowcolor{LightPurple}
	game&59	&0	&0	&0	&1	&1	&0	&0	&0	&0	&1	&1	&1	&95.13&\\
		\rowcolor{LightPurple}
	&57	&0	&1	&0	&1	&1	&0	&0	&1	&1	&1	&1	&0	&95.29 &\\
		\rowcolor{LightPurple}
	&59	&0	&1	&0	&1	&1	&0	&0	&0	&0	&1	&1	&0	&95.50 &\\
		\rowcolor{LightPurple}
	&60	&0	&1	&0	&0	&1	&0	&0	&0	&0	&1	&1	&0	&95.83 &\\ \hline \hline

    \end{tabular}
      \begin{tablenotes}
\item[1] Full board: 32 pieces
\item[2] Opening: Full board - approx 8-10 pieces
\item[3] Middle Game: Full board - approx 15-20 pieces
\item[4] End Game: 0-1 rooks, bishops, knights, 0-2 pawns, 0-1 queen.
\item[5] Expected accuracy of 5 sample calculations in each category use Figure \ref{fig:assf}.
\item[6] Actual accuracy are mean of those accuracy of those positions.
  \end{tablenotes}

    \end{tiny}
\end{table}

\subsection{Hypothesis Testing For Speed of Analysis}
Our hypothesis was that ARChessAnalyzer would be significantly faster than manual entry into a chess engine for valid outcomes. To verify our hypothesis, we setup our experiment as follows. For chess diagram entry and analysis, we chose the chess.com, which uses Stockfish as its back engine. In order to pick a variety of chess diagrams we chose five from beginning positions, mid-game positions and end game positions as follows (Table \ref{tab:chess_diagrams}). The results for manual entry and ARChessAnalyzer are tabulated in (Table \ref{tab:chess_hypothesis}). Two parameters $\mu_0$ and $\mu_a$ corresponding to the average time taken for direct entry and via the app are calculated, and so are the standard deviation ($\sigma$) (Table \ref{tab:chess_hypothesis}). Since, the number of sample size of each set is small ($n=5$) we use the student distribution t-statistic. Our $H_a$ is one tailed, and using $n-1 = 4$ degrees of freedom we see all the sets meet $t>8.610$ corresponding to $99.9995\%$. We accept the hypothesis. The highest confidence is for mid-games because the entry of the pieces are more and random. In the case of endgame, the change in number of pieces is small from an empty board, and in beginning, the change in number of pieces is small from from a full board.

\textbf{Hypothesis H1 $\mu_a \ge \mu_0$}:  Using ARChessAnalyzer is faster to analyze the chess game than using manual entries for valid outcomes.

\textbf{Null Hypothesis H0 $\mu_a < \mu_0$}: Using ARChessAnalyzer has no improvement over direct StockFish entry analysis for valid outcomes.
\begin{table}[h!]
    \centering
    \caption{Game Number, Types and Number of Pieces}
    \label{tab:chess_diagrams}
    \begin{tabular}{|c|c|c|c|c|c|c|}\hline
    \rowcolor{LightCyan}
     Chess Diagram & \multicolumn{6}{|c|}{Game Number, Types and Number of Pieces} \\ \hline
     \rowcolor{LightCyan}
     n & \multicolumn{2}{|c}{Beginning} & \multicolumn{2}{|c|}{Midgame} & \multicolumn{2}{c|}{Endgame} \\ \hline \hline
    1 & Ruy Lopez & 30(5) & Polish Immortal & 31 & Reti End Game & 4 \\ \hline
     2 & Italian & 32(6) & Stamma & 21 & Lasker’s Pin & 6 \\ \hline
     3 & Siclian Defense & 32(2) & Ponziani & 14 & Saavedra position & 4 \\ \hline
4 & French Defense & 30(6) & Abu Naim & 10 & Lucena position & 5 \\ \hline
5 & Caro Kann & 32(3) & Damiano & 11 & Philidor position & 5 \\ \hline \hline
    \end{tabular}
    \begin{tablenotes}
    \begin{tiny}
\item[1] For beginning position, both the number of pieces on the board and change in pieces from starting position are shown
\end{tiny}
  \end{tablenotes}
\end{table}

\begin{table}[h!]
\centering
    \caption{Analysis Times (in sec) and Hypothesis Testing Results}
    \label{tab:chess_hypothesis}
    \begin{tabular}{|c|c|c|c|c|c|c|}\hline
    \rowcolor{LightCyan}
     Chess Diagram & \multicolumn{3}{|c|}{Manual Entry to StockFish} & \multicolumn{3}{|c|}{ARChessAnalyzer} \\ \hline
     \rowcolor{LightCyan}
     n & Beginning & Midgame & Endgame & Beginning & Midgame & Endgame \\ \hline \hline
1 & 15.21 &	347.92 & 	15.90 & 3.20 &	3.22 &	3.40 \\ \hline
2 & 18.42 &	222.79 &	12.45 &	2.60 &	3.12 &	3.50 \\ \hline
3 & 16.80 &	107.91 & 	32.67 &	4.50 &	4.25 &	3.30 \\ \hline
4 & 17.24 &	100.87 &	15.32 &	3.50 &	2.78 &	2.20 \\ \hline
5 & 15.56 &	96.24 &	25.67 & 	2.90 &	3.23 &	3.10 \\ \hline \hline
\rowcolor{LightCyan} 
$\mu$ & 16.65 &	175.15	& 20.40 &	3.34 &	3.32 &	3.10 \\ \hline 
\rowcolor{LightCyan} 
$\sigma$ & 1.30	& 109.98 &	8.48 &	0.73 &	0.55 &	0.52 \\ \hline
\rowcolor{LightCyan}
t-value & & & & 18.23 &	311.82	& 32.99 \\ \hline \hline

    \end{tabular}
\end{table}

\newpage

\section{Conclusions And Future Work}
This paper improves the state of the art in areas of chessboard segmentation and chess piece detection using convolutional neural networks, making trade-offs in model accuracy and memory footprint for integration in an handheld device on an first of its kind iOS app. The piece detection accuracy is greater than 99\% and the accuracy of prediction pipeline is 93.45\%. The AR pipeline takes about 3-4.5sec from pointing the live view camera at the chessboard to AR Overlay. We also validated our hypothesis that ARChessAnalyzer is significantly faster at analysis than manual entry for valid outcomes. The project source code can be found at \url{https://github.com/anavmehta/ARChessAnalyzer}. The app is also available on the iOS store. 

Our hope is that the instantaneous feedback this app provides will help chess learners at all levels all over the world. 

The following are the areas of focus of future research.
\begin{itemize}
\item \textbf{Chessboard segmentation} 

The algorithm can be improved to make the camera angle more tolerant to roll, yaw, and pitch and infer hidden points and lines to detect partially hidden pieces.
\item \textbf{Chess piece detection}

More images of chessboards and pieces, from different chess sets - with variation of top and side angles to better capture the shape of the piece, will improve the robustness of classifiers to occlusions, artifacts and intra-class variations.
AlexNet was chosen based on ease of training and accuracy. A more systematic evaluation of recent research of in-device models will help further reduce memory footprint while retaining accuracy.


\item \textbf{Chess engine and game state recognition}

The state of the board sometimes requires knowledge of previous immediate moves (e.g en-passant or castling). This would require deeper analysis, and/or the user overrides.
StockFish is the most popular and was easy to integrate in iOS. However, players may want to switch to other engines based on their preference or engine strengths.
\end{itemize}

Finally, for worldwide acceptance as a learning chess app it has to be ported to Android.

\newpage

\bibliographystyle{plain}

\end{document}